\documentclass[a4paper,12pt]{article}
\usepackage[esperanto,portuguese,english]{babel}
\usepackage{epsfig}
\usepackage{float}

\newcommand{\ppparallel}[1]{} 
\newcommand{\ppmaldekstra}[1]{} 
\newcommand{\ppdekstra}[1]{}
\usepackage{parallel} \renewcommand{\ppparallel}[1]{#1} 
\newcommand{\comentario}[1]{}

\newcommand{\dd}{\mathrm{d}}

\ppparallel{    
\newlength{\pplw}\setlength{\pplw}{0.475\textwidth}
\newlength{\pprw}\setlength{\pprw}{0.485\textwidth}

\newcommand{\ppn}{\noindent}              
\newcommand{\ppl}[1]{\ParallelLText{\selectlanguage{esperanto}#1}}
\newcommand{\ppr}[1]{\ParallelRText{\selectlanguage{english}#1}\ppp}
\newcommand{\ppln}[1]
{\ParallelLText{\ppn \selectlanguage{esperanto}#1}} 
\newcommand{\pprn}[1]
{\ParallelRText{\ppn \selectlanguage{english}#1}\ppp} 

\newcommand{\ppsection}[3][0ex]{\vspace{2em} 
\ppl{\section{#2} \vspace{#1}} \ppa \nopagebreak
\ppR{\section{#3}} \ppp \nopagebreak} 

\newcommand{\bea}{\vspace{-1ex}\begin{eqnarray}}
\newcommand{\eea}{\end{eqnarray}}
} 

\ppmaldekstra{
\newcommand{\ppl}[1]{\selectlanguage{esperanto}#1}
\newcommand{\ppln}[1]{\noindent \selectlanguage{esperanto}#1}
\newcommand{\ppr}[1]{\selectlanguage{english}}
\newcommand{\pprn}[1]{\noindent \selectlanguage{english}}
\newcommand{\ppsection}[3][0ex]{\section{#2}}

\newcommand{\bea}{\begin{eqnarray}}
\newcommand{\eea}{\end{eqnarray}}
}

\ppdekstra{
\newcommand{\ppl}[1]{\selectlanguage{esperanto}}
\newcommand{\ppln}[1]{\noindent \selectlanguage{esperanto}}
\newcommand{\ppr}[1]{\selectlanguage{english}#1}
\newcommand{\pprn}[1]{\noindent \selectlanguage{english}#1}
\newcommand{\ppsection}[3][0ex]{\section{#3}}

\newcommand{\bea}{\begin{eqnarray}}
\newcommand{\eea}{\end{eqnarray}}

}

\title{{\bf La relativeca tempo -- II \ppparallel{\\ The relativistic time -- II}}}
\author{F.M. Paiva \\ 
{\small Departamento de F\'isica, Unidade Humait\'a II, Col\'egio Pedro II} \\
{\small Rua Humait\'a 80, 22261-040  Rio de Janeiro-RJ, Brasil; fmpaiva@cbpf.br} 
\vspace{.7ex} \\
A.F.F. Teixeira \\
{\small Centro Brasileiro de Pesquisas F\'isicas} \\
{\small 22290-180 Rio de Janeiro-RJ, Brasil; teixeira@cbpf.br}}

\begin{document}
\selectlanguage{esperanto}
\maketitle
\thispagestyle{empty}

\begin{abstract}\selectlanguage{esperanto}
La relativeca teorio montris ke pluraj Newtonaj ideoj pri la spacotempo estas malperfektaj. Tie \^ci ni prezentas kelkajn relativecajn konceptojn iel rilatajn al tiuj ideoj: samtempecon de eventoj kaj sinkronon de horlo\^goj (amba\u u la\u u linio en la spaca reto), gravitan Doppleran efikon, kaj voja\^gon kun reveno al estinto.    

\ppparallel{\selectlanguage{english} The theory of relativity showed that several Newtonian ideas about spacetime are imperfect. We present here some relativistic concepts related to these ideas: simultaneity of events and synchronization of clocks (both along a line in the space frame), gravitational Doppler effect, and time travel.}
\end{abstract}

\ppparallel{
\begin{Parallel}[v]{\pplw}{\pprw}
}

\ppparallel{\section*{\vspace{-2em}}\vspace{-2ex}}   




\ppsection[0.6ex]{Enkonduko}{Introduction}                                       \label{Sek1}

\ppln{Anta\u ua artiklo~\cite{reltemp1} prezentis kelkajn fizikajn efikojn montrante ke la Newtona tempo malsimilas al la tempo de speciala relativeco (SR). Tie ni diskutis pri Dopplera efiko, \^gemel-paradokso, rotacio, rigida stango, kaj konstanta propra akcelo. Por tio ni koordenatigis la spacotempon per inercia referencosistemo. Tiu sistemo ta\u ugas por fiziko sen gravito. Tamen se gravito gravas, ni bezonas uzi koordinatsistemojn pli \^generalajn, kiel tiu \^ci artiklo priskribas.}
\pprn{A former article~\cite{reltemp1} presented some physical effects showing that Newtonian time is different from special relativity (SR) time. There we discussed about Doppler effect, twin paradox, rotation, rigid rod, and constant proper acceleration. To that end we coordinatized spacetime using an inertial reference system. That system is suitable to physics without gravitation. However, if gravity is important, we need use systems of coordinates more general, as this article describes.} 

\ppl{Ni komence klarigas la koncepton de normhorlo\^go, de propra intertempo, kaj de sekundo; kaj ni emfazas la konstantecon de lumrapido en vakuo, en iu ajn gravito.}
\ppr{We initially clarify the concept of standard clock, of proper intertime, and of second; and we emphasize the constancy of light speed in vacuum, in any gravitation.}

\ppl{{\it Normhorlo\^go} estas kiel altkvalita brakhorlo\^go \cite[Sek.1]{reltemp1}. \^Ciuj normhorlo\^goj estas {\it similaj}, tio signifante ke du apudaj normhorlo\^goj montras la saman kadencon de fluo de iliaj tempoj. Tamen komence apudaj kaj sinkronaj normhorlo\^goj, poste havante malsimilajn rapidon kaj graviton, estos probable nesinkronaj en okaza renkonto. En la renkonto, la du normhorlo\^goj ankora\u u montras la saman kadencon de fluo de iliaj tempoj.}
\ppr{A {\it standard clock} is like a wrist watch of superior quality \cite[Sect.1]{reltemp1}. All standard clocks are {\it alike}, meaning that two of them, placed very close, show the same pace in the flow of their times. However, standard clocks initially neighbor and synchronous, and later submitted to different speeds and gravitation, will be probably not synchronous in an occasional reencounter. On reencounter, the two clocks still show the same pace in the flow of their times.}

\ppl{Normhorlo\^go montras la fluon de $\tau$, sia {\it propratempo}. Konvena unueco de {\it propra intertempo} $\Delta\tau$ estas la {\it sekundo}, kiu pendas nek de la movado de normhorlo\^go nek de \^gia pozicio en gravita kampo. Estas interkonsentita ke sekundo estas la intertempo de senerare 9.192.631.770 periodoj de specifita radiado el atomo de cezio. Vidu \cite[pa\^go$\!$ 18]{sek} por detaloj.}
\ppr{A standard clock shows the flow of $\tau$, its {\it propertime}. A convenient unit of {\it proper intertime} $\Delta\tau$ is the {\it second}, that depends neither on the motion of the standard clock nor on its location in a gravitational field. It is postulated that second is precisely the duration of 9.192.631.770 periods of a specified radiation from atom of cesium. See \cite[page$\!$ 18]{sek} for details.} 

\ppl{Anka\u u estas interkonsentita \cite[pa\^go$\!$ 17]{sek} ke la rapido $c$ de lumo en vakuo estas senerare $299.792.458\,m/s$. Konsekvence, {\it metro} estas moderne difinita kiel la distanco ke lumo trakuras en vakuo dum frakcio 1/299.792.458 de sekundo, senerare.} 
\ppr{It is also postulated \cite[page$\!$ 17]{sek} that the speed $c$ of light in vacuum is $299.792.458\,m/s$\, precisely. As a consequence, the {\it meter} is modernly defined as the distance that light travels in vacuum in the fraction 1/299.792.458 of a second, precisely.}

\ppl{Tiu \^ci artiklo akordas Einstein~\cite{molusco}, Landau-Lifshitz~\cite{LL}, Anderson~\cite{Anderson}, Synge~\cite{Synge2}, Misner-Thorne-Wheeler~\cite{MTW}. Tamen, malsamaj vidpunktoj ekzistas, kiel tiu de Tonkinson~\cite{Tonkinson}, ekzemple. }   
\ppr{This article goes with Einstein~\cite{molusco}, Landau-Lifshitz~\cite{LL}, Anderson~\cite{Anderson}, Synge~\cite{Synge2}, Misner-Thorne-Wheeler~\cite{MTW}. However,  different viewpoints exist, such as that of Tonkinson~\cite{Tonkinson}, for example.}

\ppsection[0.6ex]{Koordinatoj}{Coordinates}                                   \label{molusco}

\ppln{Pensu pri {\it finhava regiono} ${\cal R}^3$ de trispaco; kaverno, ekzemple, eble havante graviton. Tie ni etendas (formale) aron de dudimensiaj surfacoj, kiel grandaj kurtenoj, kaj ni orde etiketas ilin per reelaj nombroj $x^1$. Tiuj surfacoj \^generaligas la karteziajn ebenojn $x=$ konst de Euklida geometrio. Simile kiel tiuj ebenoj, surfacoj $x^1=$ konst ne tu\^sas unu la alian, kaj ilia aro plenigas regionon ${\cal R}^3$. Sed malsimile al karteziaj ebenoj, tiuj \^ci koordinatsurfacoj {\it povas malformi\^gi} la\u ulonge la tempo.}
\pprn{Consider a {\it finite region} ${\cal R}^3$ of threespace; a cavern, for example, possibly with gravity. There, we lay (formally) a set of two-dimensional surfaces, like large curtains, and orderly label them with real numbers $x^1$. These surfaces generalize the cartesian planes $x=$ const of Euclidean geometry. Like these planes, the surfaces $x^1=$ const do not touch one-another, and their set fills the entire region ${\cal R}^3$. But differently from these planes, these surfaces {\it may deform} along the time.}  

\ppl{Ni simile etendas (formale) aliajn du arojn de dudimensiaj surfacoj, kaj etiketas ilin per  reelaj nombroj $x^2=$ konst kaj $x^3$\,=\,konst. Tiuj \^ci surfacoj \^generaligas la karteziajn ebenojn $y=$ konst kaj $z=$ konst. Tiel, tridimensia {\it spaca reto} estas elektita por ${\cal R}^3$. \^Ciu punkto $P$ de la reto rilatas neambigue al triopo $x^i=[x^1, x^2, x^3]$. \^Car la reto povas malformi\^gi la\u ulonge la tempo, tial distanco inter punktoj de reto povas anka\u u varii.}   
\ppr{We similarly lay (formally) two other sets of two-dimensional surfaces, and label them with real numbers $x^2$\,=\,const and $x^3$\,=\,const. These surfaces generalize the cartesian planes $y=$ const and $z=$ const. Thus, a three-dimensional {\it spatial frame} was chosen for ${\cal R}^3$. Each point $P$ in the frame corresponds unambiguously to a triplet  $x^i=[x^1, x^2, x^3]$. Since the frame may deform along the time, the distance between points on the frame may also vary.}

\ppl{Poste, ni fiksas (formale) unu horlo\^gon ${\cal K}_P$ en \^ciu punkto $P$ de la reto; \^gi estas la {\it koordinathorlo\^go} de $P$\,, a\u u {\it loka horlo\^go} de $P$\,. \^Ciu  ${\cal K}_P$\, montras valoron {\it \^ciam pliegantan} de nedimensia variablo $t$, la {\it tempa koordinato} en $P$. ${\cal K}_P$ ne estas normhorlo\^go, ordinare. Intervalo $\Delta t$ de loka tempa koordinato estas apena\u u numero, sen unueco de intertempo (sed legu~\cite{ttau}). \^Ciu ${\cal K}_P$ havas sian kadencon de fluo de sia tempo $t$\,, kaj du koordinathorlo\^goj ordinare ne estas sinkronaj, e\^c se ili estas najbaraj. Plu, ordinare estas malkonstanta, la rilato inter la fluo de $t$ em ${\cal K}_P$ kaj la fluo de $\tau$ en normhorlo\^go fiksita \^ce $P$\,.} 
\ppr{We next fix (formally) a clock ${\cal K}_P$ in each point $P$ of the frame; it is the {\it coordinate clock} at $P$\,, or {\it local clock} at $P$\,. A ${\cal K}_P$\, shows an {\it always increasing} value of a non-dimensional variable $t$, the {\it time coordinate} at $P$. A ${\cal K}_P$ is not a standard clock, in general. An interval $\Delta t$ of local time coordinate is just a number, deprived of unity (but read~\cite{ttau}). Each ${\cal K}_P$ has its speed of flow of its time $t$\,, and two coordinate clocks usually are not synchronous, even if they are neighbor. Further, it is usually variable, the ratio between the flow of $t$ in a ${\cal K}_P$ and the flow of $\tau$ in a standard clock fixed at $P$\,.}

\ppl{Zorgo estas konsilinda, kiam ni altempigas \^ciun ${\cal K}_P$ en spaca reto: {\it najbaraj} lokaj horlo\^goj devas montri valorojn de $t$ anka\u u najbaraj. Atentu, ke voja\^ganto en spaca reto povas renkonti serion de $t$ pliegantan, a\u u sen\^san\^gantan, a\u u e\^c plietantan; Figuro~\ref{fig.FigTeTauN} ekzemplas tiun eblecon.}
\ppr{A care is worth advising, when we adjust each ${\cal K}_P$ in the spatial frame: {\it neighbor} local clocks should show also neighbor values of $t$\,. Note that a voyager in the spatial frame can find series of increasing $t$\,, or stationary, or even decreasing; Figure~\ref{fig.FigTeTauN} exemplifies that possibility.} 

\begin{figure}                                                                           
\centerline{\epsfig{file=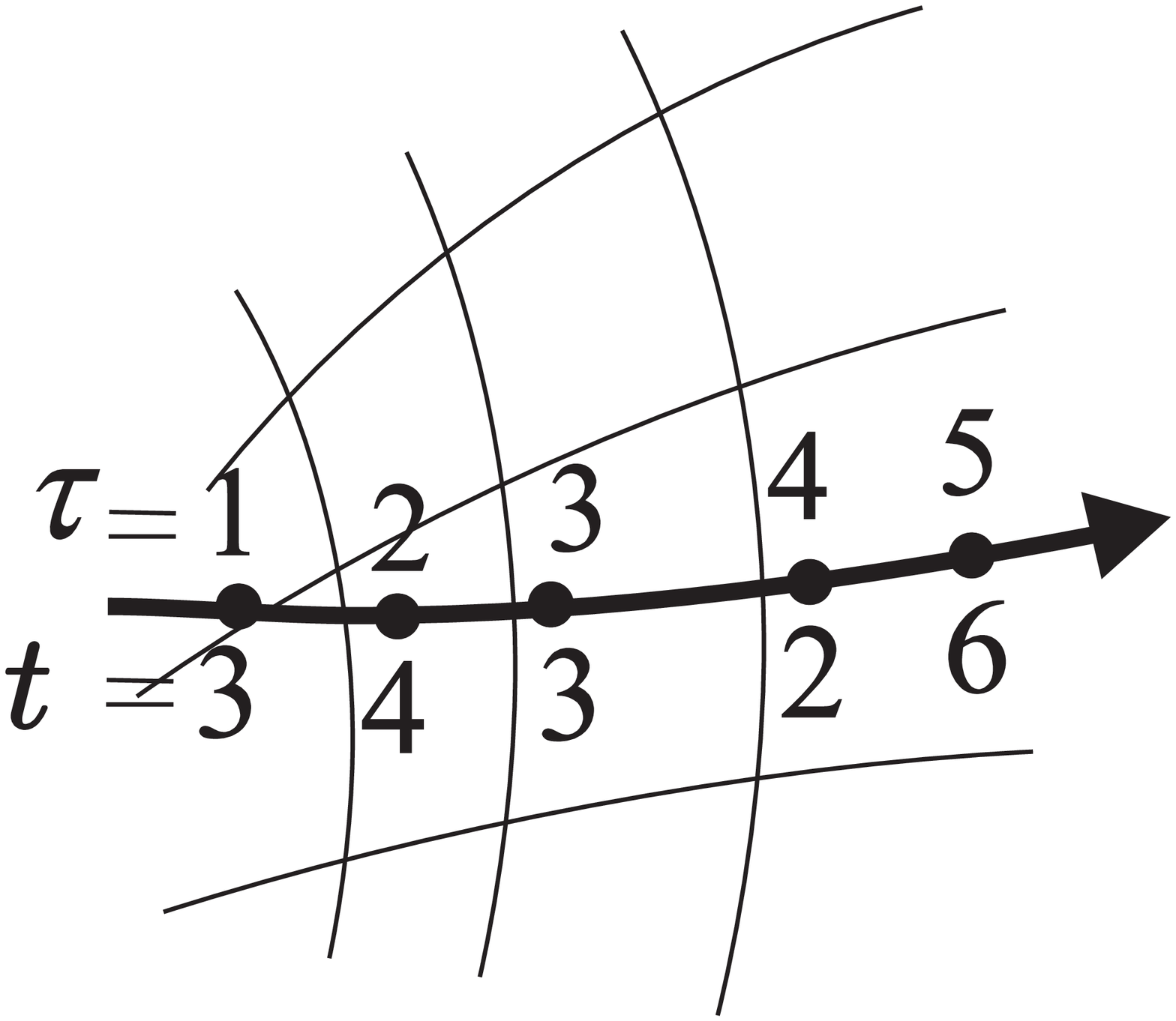,width=35mm,height=35mm}} 
\selectlanguage{esperanto}\caption{\selectlanguage{esperanto} Normhorlo\^go $\,$ movi\^gas $\,$ en $\,$ spaca reto. \^Gia propratempo \^ciam pligrandi\^gas {\mbox($\tau =\dots, 1, 2, 3, 4, 5, \dots$)}, tamen la responda loka tempa koordinato ne $\!$ \^ciam $\!$ pligrandi\^gas $\!$ ($t=\dots, 3, 4, 3, 2 , 6, \dots$).
\newline \selectlanguage{english}Figure~1: A standard clock moves in the spatial frame. Its propertime is always increasing ($\tau =\dots, 1, 2, 3, 4, 5, \dots$), although the corresponding local time coordinate not always increases ($t=\dots, 3, 4, 3, 2 , 6, \dots$).}
\label{fig.FigTeTauN} 
\end{figure}

\ppl{Einstein nomis {\it molusko}, la spacotempan koordinatsistemon konstruitan kiel tie \^ci~\cite{nome}. Kaj ni nomas {\it evento}, ion okazantan en konata punkto de spaco kaj en konata momento. En elektita molusko, evento $E$ estas specifita per kvaropo $[\,t, x^1, x^2, x^3]$\,, a\u u ekvivalente per kvaropo $x^\mu:=[x^0, x^i]$\,, estante $x^0:=ct$. Se ni elektas alian moluskon, la sama evento $E$ estos specifita per alia kvaropo $x^{\mu'}$\,. Riskante negravan konfuzon kun $t$\,, anka\u u $x^0$ estas nomita tempa koordinato de $E$.}
\ppr{Einstein called {\it molusc}, the spacetime coordinate system built as here~\cite{nome}. And we call {\it event}, something that happens in a known point of space and in a known moment. In a chosen molusc, an event $E$ is specified with a quartet $[\,t, x^1, x^2, x^3]$\,, or equivalently with the quartet $x^\mu:=[x^0, x^i]$\,, being $x^0:=ct$. If another molusc is chosen, the same event $E$ is specified with another quartet $x^{\mu'}$\,. With unimportant risk of confusion with  $t$, also $x^0$ is called time coordinate of $E$.}

\ppl{Elekti moluskon, kiel ni \^{\j}us faris, ne sufi\^cas por priskribi kelkajn fizikajn variablojn. Ekzemple, priskribo de vektoro a\u u tensoro en evento $E$ bezonas {\it vektoran bazon} en $E$\,. \^Ciu bazo ta\u ugas por specifi {\it komponojn} de vektoro a\u u tensoro. Konvenas ke bazoj en najbaraj eventoj malsimilu apena\u u infinitezime. Tie \^ci ni uzos nur la plej bonkonatajn bazojn, la {\it koordinatbazojn}, kaj iliajn dualajn~\cite[pa\^go\,80]{lich}. Tiuj bazoj estas kreitaj ekde koordinatoj, sole. Tiel, ekde nun, la vorto molusko implicos uzadon de tiuj bazoj.}
\ppr{Choosing a molusc, as we just did, is not enough to describe certain physical variables. For example, the description of a vector or tensor in an event $E$ needs a {\it vector basis} at $E$\,. A basis permits specifying the vector or tensor {\it components}. It is worth that bases in neighbor events differ only infinitesimally. Here we shall use only the best known bases, the {\it coordinate bases}, and their duals \cite[page$\!$ 80]{lich}. These bases are created from the coordinates, only. Thus, from now on, the word molusc implies use of those bases.}  

\ppl{Indas scii\^gi ke ekzistas referencosistemoj kies koordinatoj ne estas videblaj kiel tiuj de molusko de Einstein: la nulaj koordinatoj, ekzemple. Tiuj anstata\u uaj referencosistemoj simpligas kalkulojn en kelkaj specialaj fizikaj sistemoj, e\^c en speciala relativeco.}
\ppr{It is worth knowing that there are reference systems whose coordinates are not those of Einstein's molusc: the null coordinates, for example. These alternative reference systems simplify calculations in some special physical systems, even in the special relativity.}

\ppsection[0.6ex]{Metriko}{Metric}                                           \label{komponoj}

\ppln{La spacaj koordinatoj de du punktoj en spaca reto informas se tiuj punktoj estas najbaraj (infinitezime), a\u u ne. Se tamen la punktoj ne estas najbaraj, tial la koordinatoj solaj ne informas la valoron de {\it finhava} distanco inter ili; t.e., koordinatoj ne povas diri ion kiel `punkto $A$ distancas $1m$ de punkto $B$'. Simile, tempaj koordinatoj de du samlokaj eventoj (tiuj kun samaj spacaj koordinatoj) informas se la eventoj estas tempe najbaraj, a\u u ne. Sed se samlokaj eventoj ne estas najbaraj, la koordinatoj ne informas la {\it finhavan} intertempon de procezo, ion kiel `la procezo da\u uras $1s$'. Tiaj informoj bezonas uzi {\it metrikon}.}
\pprn{The spatial coordinates of two points in a spatial frame inform whether these points are neighbor (infinitesimally), or not. If however the points are not neighbor, the coordinates alone are unable to say the value of the {\it finite} separation between them; that is, the coordinates are unable to say something as `point $A$ is $1m$ apart from point $B$'. Similarly, the time coordinates of two colocal events (those with the same spatial coordinates) inform whether the events are timely neighbor, or not. But if the colocal events are not neighbor, the coordinates do not inform the {\it finite} intertime of a process, such as `the process lasts $1s$'. These informs need use a {\it metric}.} 

\ppl{Matematike, metrika kampo $g$ estas tensora, simetria, de ordo 2. Se molusko estas elektita, la komponoj de $g$ en evento $E$ estas skribitaj $g_{\nu\rho}(E)$. Metriko oferas, krom alioj, la {\it linielementon} $\dd s$ inter {\it najbaraj eventoj}. Tiel, estu $x^\mu=[ct, x^i\,]$ la koordinatoj de evento $E$, kie $t$ estas montrata per loka horlo\^go ${\cal K}_P$ de $x^i$; kaj estu $x^\mu+\dd x^\mu=[c(t+\dd t), x^i+\dd x^i\,]$ la koordinatoj de alia evento $E+\dd E$, najbara al $E$, estante $t+\dd t$ montrata per loka horlo\^go ${\cal K}_{P+\dd P}$ de $x^i+\dd x^i$. La linielemento $\dd s$ rilatantan tiujn eventojn obeas~\cite{outras}}
\ppr{Mathematically, the metric field $g$ is tensorial, symmetric, of order 2. If a molusc was built, the components of $g$ in the event $E$ are written $g_{\nu\rho}(E)$. The metric field $g$ gives, besides other things, the {\it line element} $\dd s$ between {\it neighbor events}. Thus, assume  $x^\mu=[ct, x^i\,]$ are coordinates of an event $E$, where $t$ is shown by the local clock ${\cal K}_P$ of $x^i$; and assume $x^\mu+\dd x^\mu=[c(t+\dd t), x^i+\dd x^i\,]$ are coordinates of another event $E+\dd E$, neighbor of $E$, and where $t+\dd t$ is shown by the local clock ${\cal K}_{P+\dd P}$ of $x^i+\dd x^i$. The line element $\dd s$ relating these events obeys~\cite{outras}} 

\bea                                                                                  \label{ds}
\dd s:=\sqrt{\epsilon\, g_{\mu\nu}(E)\,\dd x^\mu\,\dd x^\nu}\,, \hskip5mm \mu, \nu=0,1,2,3\,, \hspace{4mm} \epsilon=\pm1\,; 
\eea

\ppln{la signumo $\epsilon$ en (\ref{ds}) estas elektata sama al de $g_{\mu\nu}\,\dd x^\mu\,\dd x^\nu$, en \^ciu paro de eventoj.}
\pprn{the signal $\epsilon$ in (\ref{ds}) is chosen the same as that of $g_{\mu\nu}\,\dd x^\mu\,\dd x^\nu$, in each pair of events.}

\ppl{Nek $E$ nek $E+\dd E$ estas kvarvektoroj. Sed $\dd E$ estas infinitezima kvarvektoro, kun komponoj $\dd x^\mu=[c\,\dd t, \dd x^i]$\,. \^Gi estas de tempa tipo, a\u u nula tipo, a\u u spaca tipo, se $g_{\mu\nu}\,\dd x^\mu\,\dd x^\nu>0$, a\u u $=0$, a\u u $<0$, respektive.}
\ppr{Neither $E$ nor $E+\dd E$ are four-vectors. But $\dd E$ is an infinitesimal four-vector, with components $\dd x^\mu=[c\,\dd t, \dd x^i]$\,. It is timelike, or lightlike, or spacelike, according as $g_{\mu\nu}\,\dd x^\mu\,\dd x^\nu>0$, or $=0$, or $<0$, respectively.} 

\ppl{Se $\dd E$ estas de tempa tipo, tiuokaze}
\ppr{If $\dd E$ is timelike, then} 

\bea                                                                             \label{dscdtau}
\dd s=c\,\dd\tau\,, \hspace{3mm} \epsilon=+1, 
\eea 

\ppln{kie $\dd\tau>0$ estas la propra intertempo mezurata per normhorlo\^go iranta de iu evento al alia. Intervaloj de tempa tipo okazas en movado de materia korpo, kiu voja\^gas kun rapido plieta ol $c$\,.}
\pprn{where $\dd\tau>0$ is the proper intertime measured by a standard clock going from an event to the other. Timelike intervals occur in motion of a material body, that travels with speed less than $c$\,.}   

\ppl{Ekv.\,(\ref{dscdtau}) ebligas havigi komponojn $g_{\nu\rho}$ en evento $x^\mu:=[c\,t, x^i]$, per eksperimenta metodo. En tiu metodo, nur proprajn intertempojn estas mezuritaj. Ni \^{\j}etas 10 normhorlo\^gojn ekde punkto $x^i$\,, en loka momento $t$\,, en iuj ajn direktoj kaj kun iuj ajn rapidoj. Kiam la propratempo de \^ciu horlo\^go $h$ pliegi\^gas de valoro $\Delta_h\tau$ sufi\^ce malgranda, ni alnotas la tri spacajn koordinatojn $x^i+\Delta_hx^i$ de \^gia loko, kaj la montron  $t+\Delta_ht$ de loka horlo\^go. Simile kiel (\ref{ds}) kaj (\ref{dscdtau}), ni skribas la 10 ekvaciojn}
\ppr{Eq. (\ref{dscdtau}) enables obtaining the components $g_{\nu\rho}$ in an event $x^\mu:=[c\,t, x^i]$, by empirical method. In that method, only proper intertimes are measured. We throw 10 standard clocks from point $x^i$\,, in the local moment $t$\,, in random directions and with random speeds. When the propertime of each clock $h$ has advanced a value $\Delta_h\tau$ short enough, we annotate the three spatial coordinates $x^i+\Delta_hx^i$ of the place where it is, and the reading $t+\Delta_ht$ of the local clock. As in (\ref{ds}) and (\ref{dscdtau}), we write the 10 equations} 

\bea                                                                                 \label{igs}
c^2(\Delta_h\tau)^2=g_{\nu\rho}\Delta_hx^\nu\Delta_h x^\rho\,,\hspace{5mm} h=1,2, ... ,10\,, 
\eea 

\ppln{kaj solvas tiun sistemon de ne-homogenaj linearaj ekvacioj por la 10 koeficientoj  $g_{\nu\rho}$ en evento $x^\mu$.}
\pprn{and solve that system of non-homogeneous linear equations for the 10 coefficients $g_{\nu\rho}$ in event $x^\mu$.}   

\ppl{Fakte, kompono $g_{00}$ estas sciebla pli facile, en punkto $x^i$ kaj en loka momento $t$\,. En tiu momento ni fiksas normhorlo\^gon apud la loka horlo\^go de $x^i$\,. La fluo de tempo de la du horlo\^goj rilatas kiel (\ref{ds}) kaj (\ref{dscdtau}) kun $\dd x^i=0$, tial uzas nur koeficienton $g_{00}$ de metriko:}
\ppr{In fact, component $g_{00}$ can be known more easily, in point $x^i$ and local moment $t$\,. In that moment we fix a standard clock beside the local clock of $x^i$\,. The time flows of the two clocks are related through (\ref{ds}) and (\ref{dscdtau}) with $\dd x^i=0$, so uses only the coefficient $g_{00}$ of the metric:}

\bea                                                                                \label{dtau}
\dd\tau=\sqrt{g_{00}(t, x^i)}\,\dd t\,, \hskip5mm x^i={\rm konst}\,; 
\eea 

\ppln{mezurante $\dd\tau$ kaj $\dd t$\,, oni havas $g_{00}(t, x^i)$.}
\pprn{measuring $\dd\tau$ and $\dd t$\,, one has $g_{00}(t, x^i)$.}    

\ppl{Se $\dd E$ estas de spaca tipo, tial neniu korpo a\u u lumradiado povas esti en amba\u u eventoj $E$ kaj $E+\dd E$\,; tiuokaze oni difinas {\it propran distancon} inter la eventoj,}
\ppr{If $\dd E$ is spacelike, then no body or light ray can be present in both events $E$ and $E+\dd E$\,; in this case one defines the {\it proper distance} between the events,} 

\bea                                                                             \label{dlambda}
\dd\lambda:=\dd s\,, \hspace{3mm} \epsilon=-1\,; 
\eea 

\ppln{\^gi ne pendas de molusko.}
\pprn{it does not depend on the molusc.} 

\ppl{Fine, se $\dd E$ estas de nula tipo, tial $\dd s=0$\,, kaj do amba\u u propra intertempo  $\dd\tau$ kaj propra distanco $\dd\lambda$ estas nulaj. Intervaloj de nula tipo okazas en movadoj kun rapido $c$\,, kiel tiu de lumo. En laboratorio, la matematika ($\dd s=0$) kaj eksperimenta praktikeco de lumo estas konvene profitita~\cite{havigo} por malkovri la koeficientojn $g_{\mu\nu}$ en elektita molusko.}
\ppr{Finally, if $\dd E$ is lightlike, then $\dd s=0$\,, and so both proper intertime $\dd\tau$ and proper distance $\dd\lambda$ are null. Lightlike intervals occur in motions with speed $c$\,, such as light. In a laboratory, the mathematical ($\dd s=0$) and experimental practicality of light can be used~\cite{havigo} to uncover the coefficients $g_{\mu\nu}$ in a chosen molusc.}

\ppsection[0.6ex]{$\dd L$ kaj $\dd T$}{$\dd L$ and $\dd T$}                                 

\ppln{La speciala kaj la \^generala relativeco konsideras kvardimensian spacotempon. Tamen, nia homeco kutimigis nin percepti malkune la tempon kaj la spacon, en ordinaraj fenomenoj. Tio stimulas nin ser\^ci relativecajn kvantojn rilatante (tiel bone kiel estas ebla) al Newtonaj distanco kaj intertempo.}
\pprn{The special and the general relativity assume a four-dimensional spacetime. However, our humanness makes us apprehend separately space and time, in the most usual phenomena. This stimulates us to search for relativistic quantities that correspond (so well as possible) to the Newtonian distance and intertime.}  

\ppl{Propra distanco kaj propra intertempo ne estas la ser\^cataj kvantoj, \^car $\dd\lambda$ estas difinita nur se $\dd E$ estas de spaca tipo, kaj $\dd\tau$ estas difinita nur se $\dd E$ estas de tempa tipo. Ni anticipas ke la ser\^cataj kvantoj estos trovitaj en speciala inercia molusko.}
\ppr{Proper distance and proper intertime are not the quantities we are looking for, because $\dd\lambda$ is defined only if $\dd E$ is spacelike, and $\dd\tau$ is defined only if $\dd E$ is timelike. We anticipate that the quantities we look for will be found in a special inertial molusc.}
  
\ppl{Komence ni difinu distancon $\dd L$ inter {\it punktoj} $P$ kaj $P+\dd P$\,, najbaraj en spaca reto, en elektita momento. Ni procedas simile kiel en speciala relativeco.  Ni eligas lumsignalon el punkto $P=x^i$, en loka momento $t$. La signalo atingas punkton $P+\dd P=x^i+\dd x^i$ en loka momento $t+\dd_1t$, kie $\dd_1t$ (kio estas $>$, $<$,  a\u u $=0$) estas solvo de $\dd s=0$\,:}
\ppr{We start defining distance $\dd L$ between neighbor {\it points}, $P$ and $P+\dd P$\,, of a spatial frame, in a chosen moment. We proceed similarly as in the special relativity. We emit a light signal from point $P=x^i$, in local moment $t$. The signal reaches point $P+\dd P=x^i+\dd x^i$ in local moment $t+\dd_1t$, where $\dd_1t$ (that is $>$, $<$, or $=0$) is solution of $\dd s=0$\,:}  

\bea                                                                             \label{ds2nulo}
g_{00}(c\,\dd_1t)^2+2g_{0i}(c\,\dd_1t)\dd x^i+g_{ij}\dd x^i\dd x^j=0\,,  
\eea 
\vspace{-4mm}
\ppln{do}  
\pprn{so}
\vspace{-4mm}
\bea                                                                                 \label{dt1}
\dd_1t=\frac{1}{c\sqrt{g_{00}}}\left(-h_{i}\dd x^i+\epsilon_1\sqrt{h_{ij}\,\dd x^i\dd x^j}\,\right)\,,  
\eea
\vspace{-4mm}

\ppln{kie} 
\pprn{where}
\vspace{-4mm}
\bea                                                                                 \label{hos}
\epsilon_1:=\pm1\,, \hspace{3mm} h_i(x^\mu):=\frac{g_{0i}}{\sqrt{g_{00}}}\,\,, \hspace{3mm} h_{ij}(x^\mu):=\frac{g_{0i}g_{0j}}{g_{00}}-g_{ij}\,. 
\eea

\ppln{La signalo reflektas de $P+\dd P$ en loka momento $(t+\dd_1t)$, kaj revenas al $P$ en loka momento $(t+\dd_1t)+\dd_2t$, kie $\dd_2t$ (kio anka\u u estas $>$, $<$, a\u u $=0$) anka\u u estas solvo de $\dd s=0$:}
\pprn{The signal reflects from $P+\dd P$ in local moment $(t+\dd_1t)$, and returns to $P$ in local moment $(t+\dd_1t)+\dd_2t$, where $\dd_2t$ (that also is $>$, $<$, or $=0$) is also solution of $\dd s=0$:}

\bea                                                                            \label{ds2outro}
g_{00}(c\,\dd_2t)^2-2g_{0i}(c\,\dd_2t)\dd x^i+g_{ij}\dd x^i\dd x^j=0\,, 
\eea
\vspace{-4mm}
\ppln{do} 
\pprn{so}
\vspace{-2mm}
\bea                                                                                 \label{dt2}
\dd_2t=\frac{1}{c\sqrt{g_{00}}}\left(+h_{i}\dd x^i+\epsilon_2\sqrt{h_{ij}\,\dd x^i\dd x^j}\,\right), \hspace{3mm} \epsilon_2:=\pm1\,. 
\eea

\ppln{Inter la eligo de la signalo kaj la reveno, la montro de ${\cal K}_P$ pligrandi\^gas de}
\pprn{Between emission of the signal and return, the reading of ${\cal K}_P$ increases} 

\bea                                                                                \label{soma}
\dd_1t+\dd_2t= \frac{\epsilon_1+\epsilon_2}{c\sqrt{g_{00}}}\sqrt{h_{ij}\,\dd x^i\dd x^j}\,; 
\eea

\ppln{certe $\dd_1t+\dd_2t\!>\!0$, tial nur valoroj $\epsilon_1=+1$ kaj $\epsilon_2=+1$ validas en (\ref{soma}). Do, la propra intertempo $\dd\tau$ pasita en $P$\,, inter eligo de signalo kaj reveno, rilatas al (\ref{dtau}) kaj (\ref{soma}) kiel}
\pprn{surely $\dd_1t+\dd_2t\!>\!0$, so only values $\epsilon_1=+1$ and $\epsilon_2=+1$ are valid in (\ref{soma}). Thus, the proper intertime $\dd\tau$ elapsed at $P$\,, between emission and return of signal, relates to (\ref{dtau}) and (\ref{soma}) as} 

\bea                                                                               \label{dtauL}
\dd\tau=\sqrt{g_{00}}\left(\dd_1t+\dd_2t\right) =\frac{2}{c}\sqrt{h_{ij}\,\dd x^i\dd x^j}\,. 
\eea

\ppln{Ni difinas $\dd L:= c\,\dd\tau/2$\,, t.e., la duono de spaco trakurita per lumsignalo, kun rapido $c$, dum tiu propra intertempo. Uzante (\ref{dtauL}) ni fine ricevas $\dd L:=\sqrt{h_{ij}\,\dd x^i\dd x^j}$\,.}
\pprn{We define $\dd L:= c\,\dd\tau/2$\,, that is, half the space covered by the light signal, with speed $c$, during that proper intertime. Using (\ref{dtauL}) we finally have $\dd L:=\sqrt{h_{ij}\,\dd x^i\dd x^j}$\,.} 

\ppl{Resume, la {\it distanco} inter du punktoj najbaraj en la reto estas}
\ppr{In short, the {\it distance} between two neighbor points in the frame is}  

\bea                                                                                  \label{dL}
\dd L:=\sqrt{\left(\frac{g_{0i}g_{0j}}{g_{00}}-g_{ij}\right)\dd x^i\dd x^j}\,. 
\eea

\ppln{\^Car la metrikaj koeficientoj ordinare varias kun $t$\,, tial anka\u u la distancoj $\dd L$ ordinare varias kun $t$\,. Kaj tiel kiel en Euklida geometrio, $\dd L$ neniam estas negativa; kaj \^gi estas nula se nur la tri spacaj komponoj $\dd x^i$ estas nulaj, t.e., se la punktoj $P$ kaj $P+\dd P$ koincidas.}
\pprn{Since the metric coefficients usually vary with $t$\,, also the distances $\dd L$ usually vary with $t$\,. And as in Euclidean geometry, $\dd L$ is never negative; and it is null only if the three spatial components $\dd x^i$ are null, that is, if the points $P$ and $P+\dd P$ coincide.} 

\ppl{Ekvacio (\ref{dL}) difinas anka\u u {\it distancon inter najbaraj eventoj}, en elektita molusko. Atentu ke $\dd L$ ne pendas de diferenco $\dd t$ inter la tempaj koordinatoj de eventoj. Tamen $\dd L$ povas varii per \^san\^go de molusko. Tiu fakto estas jam konita en speciala relativeco, per nomo plivasti\^go (a\u u malplivasti\^go) de Lorentz.}
\ppr{Equation (\ref{dL}) defines also {\it distance between neighbor events}, in a chosen molusc. Note that $\dd L$ does not depend on the difference $\dd t$ between the time coordinates of events. However, $\dd L$ may vary under change of molusc. This is already known in SR, under denomination Lorentz dilation (or contraction).} 

\ppl{Ni povas ricevi (\ref{dL}) per alia maniero~\cite[pa\^go\,348]{Anderson}: en evento $E$ ni konstruas vektoron}
\ppr{We can get (\ref{dL}) in another way \cite[page$\!$ 348]{Anderson}: in an event $E$ we construct the vector}

\bea                                                                               \label{taumu}
\tau^\mu:=\delta^\mu_0/\sqrt{g_{00}}\,, \hspace{3mm} \tau_\mu=g_{0\mu}/\sqrt{g_{00}}\,;
\eea

\ppln{\^gi estas de tempa tipo kaj unara ($\tau^\mu\tau_\mu=1$)\,, kaj celas estonton ($\tau^0>0$). Ni disigu vektoron $\dd x^\mu$ \^ce $E$ per kompono $c\,\dd T$ en direkto $\tau^\mu$\,, kaj kompono $\dd L^\mu$ en hiperebeno orta al $\tau^\mu$\,:}
\pprn{it is timelike and unitary ($\tau^\mu\tau_\mu=1$)\,, and points to the future ($\tau^0>0$). We decompose a vector $\dd x^\mu$ anchored in $E$ into component $c\,\dd T$ in direction $\tau^\mu$\,, and component $\dd L^\mu$ in the hyperplane normal to $\tau^\mu$\,:} 

\bea                                                                              \label{decomp}
\dd x^\mu=c\,\dd T\,\tau^\mu+\dd L^\mu\,, \hspace{3mm} \tau_\mu\dd L^\mu=0\,; 
\eea

\ppln{tial $c\,\dd T=\tau_\mu\dd x^\mu$\,, kaj konsekvence}
\pprn{then $c\,\dd T=\tau_\mu\dd x^\mu$\,, and consequently}

\bea                                                                                \label{dXmu}
\dd L^\mu = \dd x^\mu-\tau^\mu\tau_\nu\dd x^\nu = 
(\delta^\mu_{\,\,\nu}-\tau^\mu\tau_\nu)\dd x^\nu\,.
\eea 

\ppln{Tiu $\dd L^\mu$ estas kvarvektoro de spaca tipo, kies normo $g_{\mu\nu}\dd L^\mu\dd L^\nu$ estas $-(\dd L)^2$. T.e.,}
\pprn{This $\dd L^\mu$ is a spacelike four-vector, whose norm $g_{\mu\nu}\dd L^\mu\dd L^\nu$ is $-(\dd L)^2$. That is,} 

\bea                                                                               \label{dLgmn}
(\dd L)^2=(g_{0\mu}g_{0\nu}/g_{00}-g_{\mu\nu})\dd x^\mu\dd x^\nu\,;
\eea

\ppln{\^car nur $\mu=i$ kaj $\nu=j$ kontribuas al (\ref{dLgmn}), tial ni rericevas (\ref{dL}).}
\pprn{since only $\mu=i$ and $\nu=j$ contribute to (\ref{dLgmn}), we reobtain (\ref{dL}).}

\ppl{Nun ni esploru $c\,\dd T:=\tau_\mu\dd x^\mu$\,, t.e.,}
\ppr{We now explore $c\,\dd T:=\tau_\mu\dd x^\mu$\,, that is,} 

\bea                                                                                  \label{dT}
\dd T:=\frac{g_{0\mu}\dd x^\mu}{c\sqrt{g_{00}}}\,. 
\eea 

\ppln{Ni diras ke $\dd T$ estas {\it intertempo} de evento $x^\mu$ al evento $x^\mu+\dd x^\mu$, en elektita molusko. \^Car $\tau^\mu$ celas estonton, tial $g_{0\mu}\dd x^\mu$ pozitiva ($\dd T>0$) implicas ke evento $x^\mu+\dd x^\mu$ estas tempe {\it posta} al evento $x^\mu$\,, en tiu molusko. Klare, $g_{0\mu}\dd x^\mu$ negativa implicas la malon. Kaj ni diras ke du najbaraj eventoj estas {\it samtempaj} se $g_{0\mu}\dd x^\mu$ estas nula ($\dd T=0$), en tiu molusko.}
\pprn{We say that $\dd T$ is the {\it intertime} from event $x^\mu$ to event $x^\mu+\dd x^\mu$, in a chosen molusc. Since $\tau^\mu$ points to the future, then a positive $g_{0\mu}\dd x^\mu$ ($\dd T>0$) implies that the event $x^\mu+\dd x^\mu$ occurs {\it after} the event $x^\mu$\,, in that molusc. Of course a negative $g_{0\mu}\dd x^\mu$ implies the opposite. And we say that two neighbor events are {\it simultaneous} if $g_{0\mu}\dd x^\mu$ is null ($\dd T=0$), in that molusc.}

\ppl{Atentu ke signumo de $\dd t$ ne estas decidiga, en tiuj difinoj. Fakte, evento $E+\dd E$\,, supozita tempe posta al evento $E$\,, povas havi $\dd t>0$ a\u u $\dd t=0$ a\u u $\dd t<0$\,. Tio estas jam sugestita en Figuro~\ref{fig.FigTeTauN}. \^Car $g_{0\mu}\dd x^\mu\equiv\dd x_0$\,, samtempeco estas pli kompakte esprimebla per $\dd x_0=0$\,.}
\ppr{Note that the signal of $\dd t$ is not deciding, in these definitions. In fact, an event $E+\dd E$\,, assumed timely after event $E$\,, can have $\dd t>0$ as well as $\dd t=0$ or $\dd t<0$\,. This is already suggested in Figure~\ref{fig.FigTeTauN}. Since $g_{0\mu}\dd x^\mu\equiv\dd x_0$\,, simultaneity can be expressed more compactly by $\dd x_0=0$\,.}
 
\ppl{Atentu anka\u u en (\ref{dT}) ke la intertempo $\dd T$ pendas de spacaj komponoj $\dd x^i$, sed ne pendas de nure spacaj metrikaj koeficientoj $g_{ij}$. Malsimile al propra intertempo $\dd\tau$, kiu estas difinita nur se $\dd x^\mu$ estas de tempa tipo, la intertempo $\dd T$ estas difinita por $\dd x^\mu$ de ia ajn tipo.}
\ppr{Also note in (\ref{dT}), that the intertime $\dd T$ depends on the spatial components $\dd x^i$, but not on the metric coefficients purely spatial, $g_{ij}$. Differently from the proper intertime $\dd\tau$, that is defined only if $\dd x^\mu$ is timelike, the intertime $\dd T$ is defined for $\dd x^\mu$ of any type.} 

\ppl{Simile kiel $\dd L$\,, anka\u u $\dd T$ pendas de molusko. Atentu ankora\u u ke}
\ppr{Similarly to $\dd L$\,, also $\dd T$ depends on the molusc. Still note that}  

\bea                                                                                \label{dsTL}
\epsilon(\dd s)^2=(c\,\dd T)^2-(\dd L)^2\,.  
\eea 

\ppln{Do, kvankam $\dd T$ kaj $\dd L$ pendas de molusko, ilia speciala kombino (\ref{dsTL}) ne pendas. La simileco de (\ref{dsTL}) kun analoga ekvacio en speciala relativeco ne estas akcidenta. Fakte, $\dd T$ kaj $\dd L$ estas intertempo kaj distanco inter najbaraj eventoj, mezuritaj en inercia referencosistemo fiksita en $x^\mu$\,.}
\pprn{So, although $\dd T$ and $\dd L$ depend on the molusc, their special combination (\ref{dsTL}) does not depend. The similarity between (\ref{dsTL}) and analogous equation in the special relativity is not fortuitous. Really, $\dd T$ and $\dd L$ are intertime and distance between neighbor events, measured in an inertial system fixed in $x^\mu$\,.}    

\ppl{Se $\dd x^\mu$ estas de {\it tempa tipo}, tial estas moluskoj kie la distanco $\dd L'$ inter la eventoj estas nula (samlokaj eventoj); tial la propra intertempo $\dd\tau$ estas la modulo de intertempo $\dd T\,'$ mezurita en iu ajn el tiuj moluskoj. Sed se $\dd x^\mu$ estas de {\it spaca tipo}, tial estas moluskoj kie la intertempo $\dd T\,'$ estas nula (samtempaj eventoj); tial la propra distanco $\dd\lambda$ estas la distanco $\dd L'$ mezurita en iu ajn el tiuj moluskoj. Fine, se $\dd x^\mu$ estas de {\it nula tipo}, tial la propra intertempo $\dd\tau$ kaj la propra distanco $\dd\lambda$ estas nulaj. Sed la intertempo $\dd T$ kaj la distanco $\dd L$ estas ne-nulaj, plue ili havas saman modulon ($|c\dd T|=\dd L$)\,, kies valoro pendas de molusko.}
\ppr{If $\dd x^\mu$ is {\it timelike}, then there are moluscs where the distance $\dd L'$ between the events is null (colocal events); then the proper intertime $\dd\tau$ is the modulus of intertime $\dd T\,'$ measured in any of these moluscs. But if $\dd x^\mu$ is {\it spacelike}, then there are moluscs where the intertime $\dd T\,'$ is null (simultaneous events); then the proper distance $\dd\lambda$ is the distance $\dd L'$ measured in any of these moluscs. Finally, if $\dd x^\mu$ is {\it lightlike}, then the proper intertime $\dd\tau$ and the proper distance $\dd\lambda$ are null. But the intertime $\dd T$ and the distance $\dd L$ are not null, further they have the same modulus ($|c\dd T|=\dd L$)\,, whose value depends on the molusc.}  

\ppsection[0.6ex]{Kelkaj rimarkoj}{Some notes}                                   \label{Le}

\ppln{En difino (\ref{dL}), atentu ke {\it \^ciuj} 10 koeficientoj $g_{\mu\nu}$ kontribuas al $\dd L$\,; kaj \^car tiuj koeficientoj ordinare varias la\u u la tempo, la distanco  $\dd L$ inter punktoj najbaraj en la reto anka\u u ordinare varias, tial la reto ordinare malformi\^gas la\u u la tempo. Atentu anka\u u ke \^car $(g_{0i}\dd x^i)(g_{0j}\dd x^j)\geq0$, tial la miksitaj komponoj $g_{0i}$ de metriko, se ne-nulaj, pligrandigas distancojn $\dd L$.}
\pprn{In definition (\ref{dL}), note that {\it all} 10 coefficients $g_{\mu\nu}$ contribute to $\dd L$\,; and since these coefficients usually vary along time, the distance $\dd L$ between neighbor poins in the frame also varies, so the spatial frame usually also deforms along time. Still note that since $(g_{0i}\dd x^i)(g_{0j}\dd x^j)\geq0$, then the mixed components $g_{0i}$ of the metric, if not null, increase the distances $\dd L$.} 

\ppl{\^Car $h_{ij}\dd x^i\dd x^j$ en (\ref{dtauL}) estas ne-negativa por iuj ajn $\dd x^i$, koeficientoj $h_{ij}$ difinitaj en (\ref{hos}) obeas la $3+3+1=\,7$ neegalecojn~\cite[pa\^go~236]{LL}}
\ppr{Since $h_{ij}\dd x^i\dd x^j$ in (\ref{dtauL}) is non-negative for any $\dd x^i$, the coefficients $h_{ij}$ defined in (\ref{hos}) obey the $3+3+1=\,7$ inequalities~\cite[page$\!$ 236]{LL}} 

\bea                                                                             \label{formhij}
h_{ii}>0\,, \hspace{3mm} h_{ii}h_{jj}>(h_{ij})^2\,, \hspace{3mm} {\rm det}(h_{ij})>0\,. 
\eea

\ppln{Tiuj neegalecoj faras ke $\dd L$ estu nula nur se la tri komponoj $\dd x^i$ estas nulaj. Kaj uzante $h_{ij}$ de (\ref{hos}), kaj (\ref{formhij}), ni vidas ke koeficientoj $g_{\mu\nu}$ obeas la  $1+3+3+1=8$ neegalecojn}
\pprn{These inequalities make $\dd L$ null only if the three components $\dd x^i$ are null. And using $h_{ij}$ from (\ref{hos}), and (\ref{formhij}), we see that the coefficients $g_{\mu\nu}$ obey the $1+3+3+1=8$ inequalities} 

\bea \nonumber                                          
g_{00}>0\,, \hspace{5mm} g_{00}g_{ii}<(g_{0i})^2\,,
\eea
\vspace{-11mm}
\bea                                                                               \label{desig}
{} 
\eea 
\vspace{-12mm} 
\bea \nonumber                                          
g_{00}(g_{ii}g_{jj}-g_{ij}^2)-g_{0i}(g_{0i}g_{jj}-g_{ij}g_{0j})-g_{0j}(g_{0j}g_{ii}-g_{0i}g_{ij})>0\,,\hspace{5mm} {\rm det}(g_{\mu\nu})<0\,.
\eea

\ppl{Pro {\it arbitra} \^san\^go de molusko, kvantoj $h_{ij}$\,, $g_{\mu\nu}$\,, kaj $\dd x^\mu$ ordinare varias, en eventoj $E$ kaj $E+\dd E$. Ni povas demandi se la specialaj kombinoj (\ref{dL}) kaj (\ref{dT}) de tiuj kvantoj lasas ke $\dd L$ kaj $\dd T$ estu ne-variantaj. Kiel en speciala relativeco, respondo estas {\it ne}, ordinare.}
\ppr{Under {\it arbitrary} change of molusc, the quantities $h_{ij}$\,, $g_{\mu\nu}$\,, and $\dd x^\mu$ usually vary, in events $E$ and $E+\dd E$. We may ask whether the special combinations (\ref{dL}) and (\ref{dT}) of these quantities leave $\dd L$ and $\dd T$ invariant. As in the special relativity, the answer is {\it no}, in general.} 

\ppl{Kontra\u ue, estas {\it specialaj} \^san\^goj de molusko $[t, x^i]\rightarrow[t', x^{i'}]$ kiuj lasas ne-variantaj {\it \^ciun} infiniteziman distancon $\dd L$ kaj {\it \^ciun} infiniteziman intertempon $\dd T$\,. En tiuj \^san\^goj de molusko, la spaciaj koordinatoj transformas kiel $x^i\rightarrow x^{i\,'}(x^j)$\,. T.e., la \^san\^go de iu spaca reto al la alia ne pendas de tempo. La responda tempa transformo povas esti \^generala, $t\rightarrow t'(t, x^i)$. Atentu ke la Lorentzaj transformoj por la spaca reto, uzataj en speciala relativeco, pendas de $t$\,; ekzemple, $x\rightarrow x'=\gamma(x-vt/c)$. Bonkonate, tiuj transformoj malplivastigas la spacan reton.}
\ppr{On the contrary, there are {\it special} changes of molusc $[t, x^i]\rightarrow[t', x^{i'}]$ that leave unchanged {\it every} infinitesimal distance $\dd L$ and {\it every} infinitesimal intertime $\dd T$\,. In these changes of molusc, the spatial coordinates transform as $x^i\rightarrow x^{i\,'}(x^j)$\,. That is, the change from one spatial frame to the other does not depend on time. The corresponding time transformation can be general, $t\rightarrow t'(t, x^i)$. Note that the Lorentz transforms for the spatial frame, used in special relativity, depend on $t$\,; for example, $x\rightarrow x'=\gamma(x-vt/c)$. As is well known, these transforms contract the spatial frame.} 

\ppl{Ni vidis ke $\dd L:=\sqrt{h_{ij}\,\dd x^i\dd x^j}$ estas distanco inter punktoj najbaraj en spaca reto de elektita molusko. Se punktoj ne estas najbaraj, la distanco inter ili estas difinita per integralo de $\dd L$ la\u u la geodezio unuiganta ilin, uzante $h_{ij}$ kiel tridimensia metriko. \^Car tiuj koeficientoj ordinare varias la\u u la tempo, anka\u u la distanco ordinare varias.}
\ppr{We saw that $\dd L:=\sqrt{h_{ij}\,\dd x^i\dd x^j}$ is distance between neighbor points in the spatial frame of a chosen molusc. If the points are not neighbor, the distance is defined as the integral of $\dd L$ along the geodesic connecting them, using $h_{ij}$ as components of three-dimensional metric. Since these coefficients usually vary with time, also the separations usually vary.}

\ppsection[0.6ex]{Rapido}{Velocity}                                             \label{rapido}

\ppln{Estu du najbaraj eventoj, kun koordinatoj $x^\mu$ kaj $x^\mu+\dd x^\mu$ en iu molusko. Uzante $\dd L$ kaj $\dd T$\,, ni difinas {\it rapidon} rilatan al kvarvektoro $\dd x^\mu$ per}
\pprn{Consider two neighbor events, with coordinates $x^\mu$ and $x^\mu+\dd x^\mu$ in some molusc. Using $\dd L$ and $\dd T$\,, we define a {\it velocity} associated to the four-vector $\dd x^\mu$ by}  

\bea                                                                              \label{tambem}
v:=\frac{\dd L}{\dd T}\,. 
\eea 

\ppln{Klare, $v$ pendas de molusko. \^Car $\dd L$ estas pozitiva, kaj $\dd T$ povas esti a\u u pozitiva, a\u u nula, a\u u negativa, tial rapido (\ref{tambem}) povas havi iun ajn valoron, $-\infty\leq v\leq\infty$\,. Okazas $v>0$ se $\dd T>0$\,, t.e. se evento $x^\mu+\dd x^\mu$ estas tempe posta evento $x^\mu$ (tial $g_{0\mu}\dd x^\mu>0$\,), kaj okazas $v<0$ se kontra\u ue (tial $g_{0\mu}\dd x^\mu<0$\,). Kaj okazas $|v|=\infty$ se la eventoj estas samtempaj ($\dd T=0$).}
\pprn{Of course $v$ depends on the molusc. Since $\dd L$ is positive, while $\dd T$ can be positive, or null, or negative, the velocity (\ref{tambem}) can have any value, $-\infty\leq v\leq\infty$\,. It occurs $v>0$ if $\dd T>0$\,, that is, if event $x^\mu+\dd x^\mu$ is later than event $x^\mu$ (then $g_{0\mu}\dd x^\mu>0$\,), and it occurs $v<0$ in the opposite case (then $g_{0\mu}\dd x^\mu<0$\,). And it occurs $|v|=\infty$ if the events are simultaneous ($\dd T=0$).} 

\ppl{Se $c\,\dd T>\dd L$, tial $0<v<c$, indikante ke objekto povas ekiri de punkto $x^i$ de spaca reto en loka momento $t$ kaj atingi punkton $x^i+\dd x^i$ en loka momento $t+\dd t$. Kaj se $c|\dd T|>\dd L$ sed $\dd T<0$, tial $-c<v<0$, indikante ke objekto povas iri ekde $x^i+\dd x^i$ en loka momento $t+\dd t$ kaj atingi punkton $x^i$ en loka momento $t$\,. En amba\u u okazoj $\dd t$ povas havi iun ajn signumon. Plue, (\ref{dscdtau}) kaj (\ref{dsTL}) kaj (\ref{tambem}) oferas je}
\ppr{If $c\,\dd T>\dd L$ then $0<v<c$, indicating that an object can start from point $x^i$ in the spatial frame in local moment $t$ and reach point $x^i+\dd x^i$ in local moment $t+\dd t$. And if $c|\dd T|>\dd L$ but $\dd T<0$ then $-c<v<0$, indicating that the object can start from $x^i+\dd x^i$ in local moment $t+\dd t$ and reach point $x^i$ in local moment $t$\,. In both cases $\dd t$ can have any signal. Further, (\ref{dscdtau}), (\ref{dsTL}) and (\ref{tambem}) give}   

\bea                                                                              \label{dtaudT}
\dd\tau=\sqrt{1-v^2/c^2}\,|\dd T\,|,
\eea 

\ppln{estante $v$ la rapido de objekto en tiu molusko.}
\pprn{where $v$ is speed of the object in that molusc.} 

\ppl{Se $c|\dd T|=\dd L$, tial $\dd s=0$, do lumsignalo povas voja^gi inter eventoj $x^\mu$ kaj $x^\mu+\dd x^\mu$. La\u u (\ref{tambem}), la rapido de signalo havas modulon $c$\,. La signalo iras de $x^i$ al $x^i+\dd x^i$ se $g_{0\mu}\dd x^\mu>0$, kaj en la mala direkto se $g_{0\mu}\dd x^\mu<0$.}
\ppr{If $c|\dd T|=\dd L$, then $\dd s=0$, so a light signal can connect the events $x^\mu$ and $x^\mu+\dd x^\mu$. According to (\ref{tambem}), the velocity of the signal has modulus $c$\,. The signal goes from $x^i$ to $x^i+\dd x^i$ if $g_{0\mu}\dd x^\mu>0$, and in the opposite direction if $g_{0\mu}\dd x^\mu<0$.}

\ppl{Fine, se $c|\dd T|<\dd L$ tial (\ref{tambem}) indikas ke $|v|>c$\,: neniu objekto kaj neniu lumsignalo povas voja\^gi inter la eventoj. Speciale, se $\dd T=0$ kun $\dd L\neq0$, tial  (\ref{tambem}) indikas $|v|=\infty$\,.}
\ppr{Finally, if $c|\dd T|<\dd L$ then (\ref{tambem}) says that $|v|>c$\,: no object and no light signal can connect the events. In particular, if $\dd T=0$ with $\dd L\neq0$ then (\ref{tambem}) says that $|v|=\infty$\,.}

\ppsection[0.6ex]{En kurbo}{In a curve}                                        \label{kurbo}

\ppln{{\it Kurbo en spacotempo} estas rompebla je pecoj. Peco esta nomita {\it de tempa tipo} (a\u u de nula tipo, a\u u de spaca tipo) se \^ciu infinitezima intervalo de tiu peco estas de tempa tipo  (a\u u nula, a\u u spaca, respektive). Peco de tempa tipo kaj de nula tipo estas direkteblaj la\u u plieganta $T$\,, kaj tiu direkto ne varias per \^san\^go de molusko. Anka\u u peco de spaca tipo estas tiel direktebla, sed direkto povas varii per \^san\^go de molusko.}
\pprn{A {\it curve in the spacetime} can be broken into pieces. A piece is said {\it timelike} (or lightlike or spacelike) if every infinitesimal interval of it is timelike (or lightlike or spacelike, respectively). The timelike and the lightlike pieces can be oriented according to increasing $T$, and that orientation does not change under a change of molusc. Also a spacelike piece can be so oriented, but the orientation can change under a change of molusc.}

\ppl{Estu du eventoj $E_1$ kaj $E_2$, finie apartitaj. En Newtona mekaniko, la tempa apartigo inter ili estas simple $t_2-t_1$\,, kaj la spaca apartigo estas simple $|{\vec x_2}-{\vec x_1}|$\,. En speciala relativeco la temo estas pli subtila, \^car amba\u u apartigoj pendas de la inercia referencosistemo ke oni uzas. Kaj en molusko de Einstein la temo estas ankora\u u pli subtila.}
\ppr{Consider two events $E_1$ and $E_2$, finitely apart. In Newtonian mechanics the time separation between them is simply $t_2-t_1$\,, and the spatial separation is simply $|{\vec x_2}-{\vec x_1}|$\,. In special relativity the subject is more subtle, because both separations depend on the inertial reference system under use. And in a molusc of Einstein the subject is still more subtle.} 

\ppl{Finia apartigo estas sumigo de infinitezimaj apartigoj. Ni jam difinis du malsamajn infinitezimajn tempajn apartigojn: propran intertempon $\dd\tau$ kaj intertempon $\dd T$. Kaj ni difinis anka\u u du malsamajn infinitezimajn spacajn apartigojn: propran distancon $\dd\lambda$ kaj distancon $\dd L$. Do ni havas 4 malsimilajn finiajn apartigojn por konsideri. Tiuj apartigoj estos difinataj la\u u iu elektita kurbo en spacotempo. Atentu ke estas nefinia nombro de kurboj de spaca tipo kunigante du iujn ajn eventojn. Sed eble ne ekzistas kurbon de nula tipo, a\u u de tempa tipo, por fari tion.}
\ppr{Finite separation is a sum of infinitesimal separations. We already defined two different infinitesimal time separations: proper intertime $\dd\tau$ and intertime $\dd T$. And we also defined two different infinitesimal space separations: proper distance $\dd\lambda$ and distance $\dd L$. So we have 4 different finite separations to describe. These separations are to be defined along some chosen curve in spacetime. Note that there is an infinite number of spacelike curves to connect any two events. But possibly does not exist lightlike or timelike curve to do that.}   

\ppl{Se kurbo de tempa tipo komencas en evento $E_1$ kaj fini\^gas en evento $E_2$, tial la integra\^{\j}o}
\ppr{If a timelike curve starts in event $E_1$ and ends in event $E_2$, then the integral} 

\bea                                                                                \label{Dtau}
\Delta\tau = \int_{E_1}^{E_2}\dd\tau\,, 
\eea 

\ppln{kalkulita la\u u la kurbo, estas la propra intertempo inter la eventoj, la\u u tiu kurbo. Klare $\Delta\tau$ ne varias per \^san\^go de molusko. Se alia kurbo de tempa tipo estas elektita inter tiuj eventoj, la nova propra intertempo havos malsaman valoron, ordinare. Ekzisto de kurbo de tempa tipo kunigante du eventoj implicas ekzisto de nefinia nombro de kurboj de nula tipo, kaj de spaca tipo, kunigante tiujn eventojn.}
\pprn{calculated along the curve, is the proper intertime between the events, along that curve. Clearly $\Delta\tau$ is invariant under change of molusc. If another timelike curve is chosen between these events, the new proper intertime will have another value, usually. The existence of a timelike curve between two events implies existence of an infinite number of lightlike curves, and of spacelike curves, connecting these events.}   

\ppl{Se kurbo de spaca tipo estas elektita por kunigi $E_1$ kun $E_2$, tial la integra\^{\j}o}
\ppr{If a spacelike curve is chosen to connect $E_1$ with $E_2$, then the integral}  

\bea                                                                             \label{Dlambda}
\Delta\lambda = \int_{E_1}^{E_2}\dd\lambda\,, 
\eea 

\ppln{kalkulita la\u u la kurbo, estas la propralongo de kurbo inter la eventoj, la\u u tiu kurbo. Klare,  $\Delta\lambda$ ne varias per \^san\^go de molusko. Se alia kurbo de spaca tipo estas elektita inter tiuj eventoj, la propralongo de nova kurbo havos alian valoron, ordinare.} 
\pprn{calculated along the curve, is the properlength of the curve between these events, along that curve. Clearly $\Delta\lambda$ is invariant under change of molusc. If another spacelike curve is chosen between these events, the properlength of the new curve will have another value, usually.}

\ppl{Kompletante, ni difinas intertempon $\Delta T$ kaj distancon $\Delta L$\,, inter du eventoj en elektita kurbo. Amba\u u pendas de molusko. Por ia ajn kurbo kunigante $E_1$ kun $E_2$, ili estas difinitaj per}
\ppr{To complete, we define intertime $\Delta T$ and distance $\Delta L$ between two events in a chosen curve. Both are molusc dependent. For a curve of any type connecting $E_1$ and $E_2$, they are defined by} 

\bea                                                                                \label{DtDl}
\Delta T = \int_{E_1}^{E_2}\dd T\,, \hspace{5mm}  \Delta L = \int_{E_1}^{E_2}\dd L\,,
\eea  

\ppln{kie la integro estas farita la\u u la kurbo~\cite{tipos}. Speciala kurbo por integro de amba\u u (\ref{DtDl}a) kaj (\ref{DtDl}b) estas {\it spacotempa geodezio} entenanta $E_1$ kaj $E_2$. Tiu geodezio povas esti de ia tipo. Por integro de (\ref{DtDl}b), alia speciala kurbo estas geodezio entenanta punktojn $P_1$ kaj $P_2$, kalkulita kun metrikaj koeficientoj $h_{ij}$ skribitaj en (\ref{hos}c).}
\pprn{where the integration is performed along the curve~\cite{tipos}. A special curve for both integrals (\ref{DtDl}a) and (\ref{DtDl}b) is a {\it spacetime geodesic} containing $E_1$ and $E_2$. That geodesic can be of any type. To integrate (\ref{DtDl}b), another special curve is the geodesic containing points $P_1$ and $P_2$, calculated with the metric coefficients $h_{ij}$ written in (\ref{hos}c).} 

\ppl{Estas moluskoj kies intertempoj $\Delta T$ estas haveblaj sen bezoni integri (\ref{DtDl}a). Ekzemple, molusko kies kvar kvocientoj $g_{0\mu}/\sqrt{g_{00}}\,\,$ estas komponoj de kvargradiento de iu funkcio, kiu ni nomas $cT(x^\nu)$\,. Tial (\ref{dT}) diras ke la intertempo $\Delta T$ de evento $E_1$ al evento $E_2$ valoras $T(E_2)-T(E_1)$ por ia ajn kurbo, en tiu molusko.}
\ppr{There are moluscs whose intertimes $\Delta T$ can be obtained without integrating (\ref{DtDl}a). For example, a molusc whose four quocients $g_{0\mu}/\sqrt{g_{00}}\,\,$ are components of four-gradient of some function, say $cT(x^\nu)$\,. Then (\ref{dT}) says that the intertime $\Delta T$ from event $E_1$ to event $E_2$ is $T(E_2)-T(E_1)$ for any curve, in that molusc.}  

\ppl{Oni povas montri~\cite[pa\^go\,237]{LL} ke iu ajn finia peco de spacotempo permesas moluskojn kun $g_{00}=1$ kaj $g_{0i}=0$. En tiuj moluskoj, (\ref{dT}) simpli\^gas al $\dd T=\dd t$, do iu intertempo $\Delta T$ koincidas kun $\Delta t$\,.}
\ppr{One can show~\cite[page\,$\!$ 237]{LL} that any finite portion of spacetime permits moluscs with $g_{00}=1$ and $g_{0i}=0$. In these moluscs, (\ref{dT}) simplifies to $\dd T=\dd t$, so every intertime $\Delta T$ coincides with $\Delta t$\,.}  

\ppsection[0.6ex]{Samtempaj eventoj}{Simultaneous events}                      \label{simul}

\begin{figure}                                                                            
\centerline{\epsfig{file=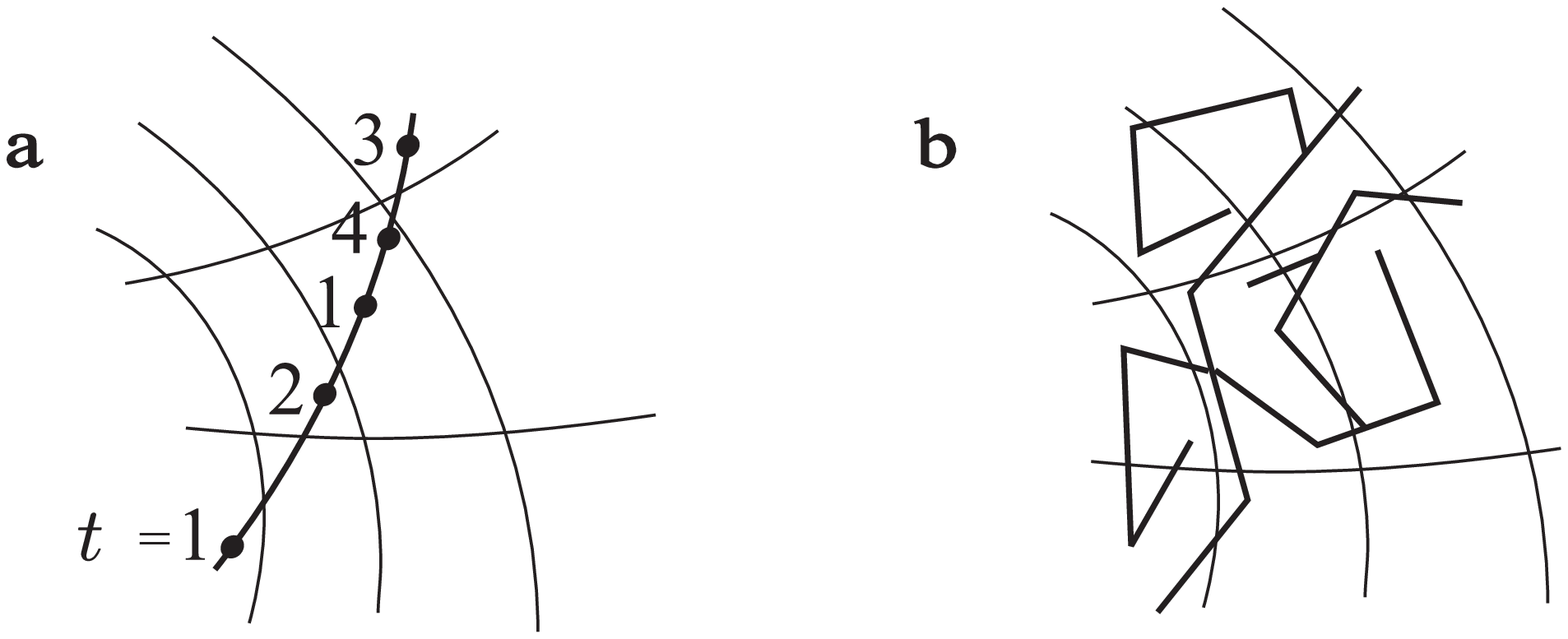,width=9cm}} 
\selectlanguage{esperanto}\caption{\selectlanguage{esperanto} {\bf a}: En malfermita linio en spaca reto, ni elektas {\it eventojn samtempajn la\u u linio}; valoro de loka tempa koordinato $t$ de eventoj varias, ordinare. {\bf b}: Arbo en spaca reto, kun iom ajn da bran\^coj, sed sen fermita spaca cirkvito; en tiu arbo ni povas elekti eventojn samtempajn la\u u linio.  
\newline \selectlanguage{english}Figure~2: {\bf a}: In an open line in the spatial frame we select {\it events simultaneous along a line}; the value of local time coordinate $t$ of these events varies from point to point, usually. {\bf b}: A tree in the spatial frame, with any quantity of branches, but without closed spatial circuit; along that tree we can select events simultaneous.}
\label{fig.FigSimulN} 
\end{figure}

\ppln{Sekcio 4 difinis samtempajn {\it najbarajn} eventojn en elektita molusko, tiel ke iliaj intertempo $\dd T$ estas nula, do~\cite[pa\^go\,237]{LL}}
\pprn{Section 4 defined simultaneous {\it neighbor}\, events in a chosen molusc, such that their intertime $\dd T$ is null, so~\cite[page$\!$ 237]{LL}}

\bea                                                                                 \label{dx0}
c\,\dd t=-\frac{g_{0i}(x^\mu)\,\dd x^i}{g_{00}(x^\mu)}\,,   
\eea

\ppln{estante $x^\mu$ kaj $x^\mu+\dd x^\mu$ koordinatoj de eventoj en molusko. Kaj ni vidis ke \^san\^go de molusko ordinare malfaras fruajn samtempecojn, sed kreas novajn.}
\pprn{where $x^\mu$ and $x^\mu+\dd x^\mu$ are components of the events in the molusc. And we saw that a change of molusc usually undoes previous simultaneities, but creates new.} 

\ppl{Nun ni difinas samtempecon de eventoj {\it finie} apartitaj en spaca reto de molusko. Por tio, ni kreas novan koncepton, tiun de {\it samtempeco la\u u linio} en spaca reto~\cite[pa\^go\,237]{LL}. Ni elektas iun ajn {\it linion} ${\cal L}$ sen memcruci\^go en spaca reto, kaj elektas iun ajn eventon $E$ en tiu lineo. Tial ni povas trovi, ekde evento $E$ kaj en amba\u u direktoj de lineo ${\cal L}$, sinsekvajn samtempajn najbarajn eventojn ($\dd T=0$) en ${\cal L}$. Kvankam samtempaj najbaraj eventoj havas valorojn de loka tempo $t$ tre similaj, eventoj en punktoj sufi\^ce apartitaj en ${\cal L}$ povas havi valorojn de $t$ {\it tre malsimilaj}. Tamen ni diras ke tiuj eventoj estas {\it samtempaj la\u u ${\cal L}$}\,. Vidu  Figuron~\ref{fig.FigSimulN}a. Tiu koncepto validas anka\u u por  {\it arbo de samtempecoj} en spaca reto de molusko, kiel en Figuro~\ref{fig.FigSimulN}b.}
\ppr{We now define simultaneity of events {\it finitely} separated in the spatial frame of a molusc. For that, a new concept is built, that of {\it simultaneity along a line} in the spatial frame~\cite[page\,$\!$ 237]{LL}. We choose any line ${\cal L}$ without self-intersection in the spatial frame, and we choose any event $E$ in that line. We then specify sequentially, starting from event $E$ and in both directions of line ${\cal L}$, simultaneous neighbor events ($\dd T=0$) along ${\cal L}$. Although simultaneous neighbor events have values of local time $t$ very similar, events in points sufficiently apart on ${\cal L}$ can have {\it very different} values of $t$\,. Even though, we shall say that all these events are {\it simultaneous along ${\cal L}$}\,. See Figure~\ref{fig.FigSimulN}a. This concept can be extended to a {\it tree of simultaneities} in the spatial frame, as in Figure~\ref{fig.FigSimulN}b.}

\ppl{Atentu ke du eventoj samtempaj la\u u linio ${\cal L}_a$ en elektita molusko ne estas samtempaj la\u u alia linio ${\cal L}_b$, ordinare. Tio ordinare malpermesas ekziston de {\it fermita} linio en spaca reto, konsistanta nure el paroj de samtempaj najbaraj eventoj. Konsekvence ne ekzistas, ordinare, eventoj finie apartitaj en spaca reto, estante samtempaj la\u u iu ajn linio. Vidu Figuron~\ref{fig.FigLaLbN}.}
\ppr{Note that two events simultaneous along a line ${\cal L}_a$ in a given molusc are not simultaneous along another line ${\cal L}_b$, in general. This prevents existence, in general, of a {\it closed} line in the spatial frame, where every pair of neighbor events is simultaneous. Consequently there do not exist, in general, events finitely separated in the spatial frame, that be simultaneous along any line. See Figure~\ref{fig.FigLaLbN}.}

\ppl{Bonkonate, \^ciu finia peco de spacotempo permesas moluskojn tiel ke \^ciuj eventoj estas samtempaj la\u u iu ajn linio~\cite[pa\^go\,$\!$ 286]{LL}. Tial ni diras ke tiuj eventoj estas {\it samtempaj} (sen indiki linion) en tiu molusko. Ekzemple, en molusko kun $g_{0i}=0$ \^ciuj eventoj kun sama valoro de $t$ estas samtempaj, la\u u (\ref{dx0}).}
\ppr{It is well known that every finite portion of spacetime permits moluscs such that all events are simultaneous along any line~\cite[page\,$\!$ 286]{LL}. We then say that these events are {\it simultaneous} (without reference to line) in that molusc. For example, in molusc with $g_{0i}=0$ all events with same value of $t$ are simultaneous, according to (\ref{dx0}).}

\ppl{Aliaj moluskoj oferantaj (tutan, por emfazo) samtempecon havas la tri kvocientojn $g_{0i}/g_{00}$ ne pendantaj de $t$, kaj plue estantaj komponoj de trigradiento de funkcio, kiun ni nomas $-ct(x^j)$. Tio estas,}
\ppr{Other moluscs that offer global simultaneity have the three cocients $g_{0i}/g_{00}$ not depending on $t$, and further being components of the three-gradient of some function, which we call $-ct(x^j)$. That is,} 

\bea                                                                           \label{especial1}
\frac{g_{0i}\,(t, x^j)}{g_{00}\,(t, x^j)}=-c\frac{\partial t(x^j)}{\partial x^i}\,. 
\eea

\ppln{Tial, kondi\^co por samtempeco (\ref{dx0}) simpli\^gas al $\dd t=\dd t(x^j)$, kies solvo estas $t-t_0=t(x^j)-t(x^j_0)$. Per vortoj, se oni elektas eventon $E_0:=[t_0, x^j_0]$\,, tial funkcio $t(x^j)$ indikas valoron de $t$ en evento $[t, x^j]$\,, kiu estas samtempa al evento $E_0$ la\u u iu ajn linio en spaca reto. Ni \^ciam povas konstrui tiel molusko, en finia peco de spacotempo.}
\pprn{So the condition for simultaneity (\ref{dx0}) reduces to $\dd t=\dd t(x^j)$, whose solution is $t-t_0=t(x^j)-t(x^j_0)$. In words, if an event $E_0:=[t_0, x^j_0]$ is chosen, the function $t(x^j)$ indicates the value of $t$ in event $[t, x^j]$\,, that is simultaneous to event $E_0$ along any line in the spatial frame. We can always build one such molusc, in a finite portion of spacetime.}

\ppl{Ni scias ke du najbaraj eventoj generas \^guste unu geodezion. Se tiuj eventoj estas samtempaj  ($\dd T=0$) en iu molusko, tial la geodezio estas de spaca tipo. Kaj la rapido (\ref{tambem}) rilata al eventoj estas nefinia en tiu molusko. Ni difinas {\it geodezio de samtempecoj}, geodezion de spaca tipo kie \^ciuj najbaraj eventoj estas samtempaj, en tiu molusko~\cite{GeSiSchw,SomRay}. Malsimile al aliaj geodezioj, \^ci tia ne estas direktebla per plieganta $T$\,. En Newtona interpreto, \^gi prezentas movadon de hipoteza objekto sentiva al gravito, sed voja\^ganta kun nefinia rapido.}
\ppr{We know that two neighbor events generate just one geodesic. If they are simultaneous ($\dd T=0$) in some molusc, then the geodesic is spacelike. And the velocity (\ref{tambem}) associated to the two events is infinite in that molusc. We define as {\it geodesic of simultaneities} the spacelike geodesic whose (all) neighbor events are simultaneous, in that molusc~\cite{GeSiSchw,SomRay}. Differently from other geodesics, it can not be oriented with increasing $T$\,. In a Newtonian interpretation, it represents the motion of a hipothetical object sensitive to gravitation, but traveling with infinite velocity.} 

\begin{figure}                                                                            
\centerline{\epsfig{file=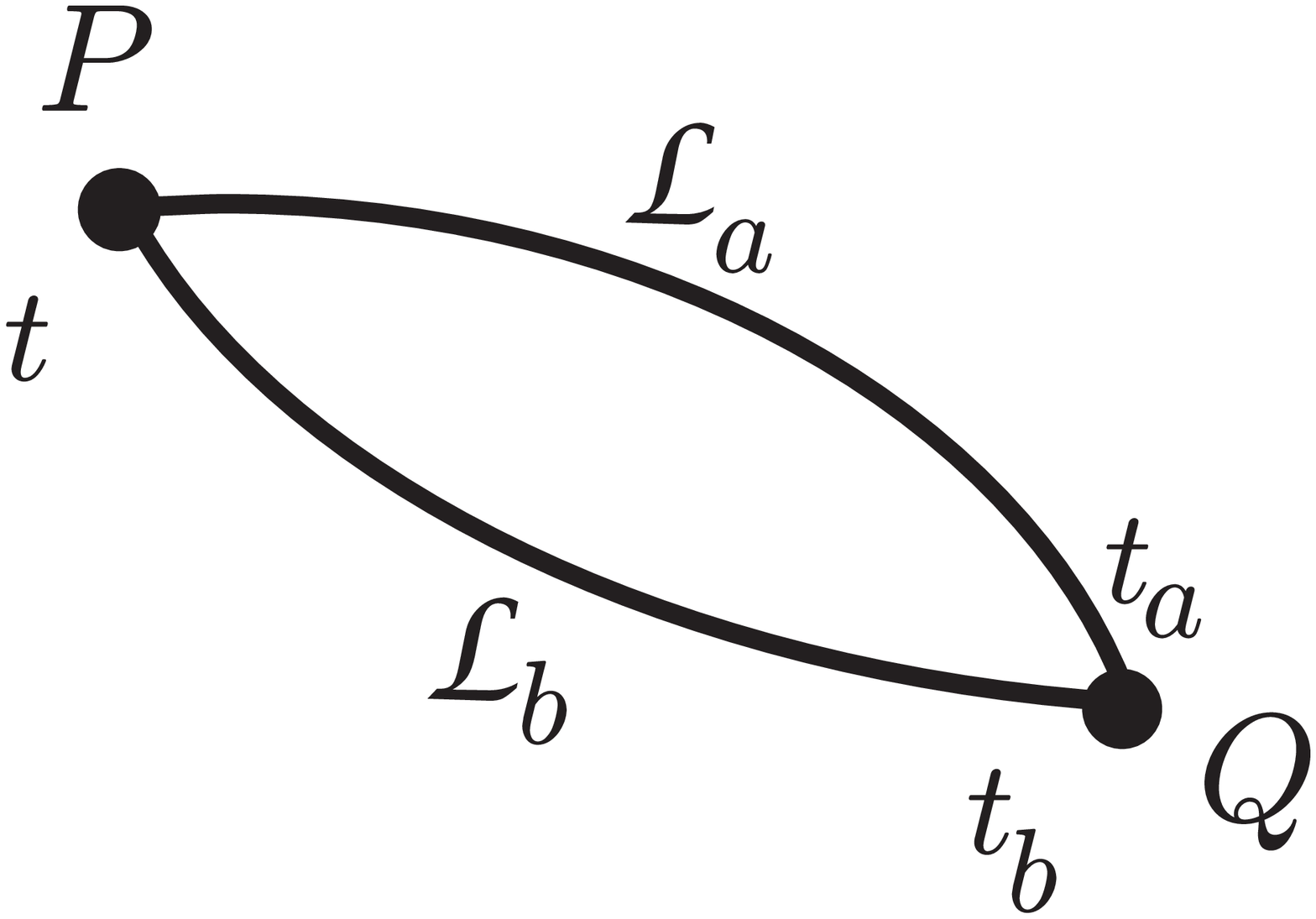,width=35mm}} 
\selectlanguage{esperanto}\caption{\selectlanguage{esperanto} $P$ kaj $Q$ estas punktoj de spaca reto, kunigitaj per linioj ${\cal L}_a$ kaj ${\cal L}_b$ en la reto. Supozu ke evento $[t_a, Q]$ estas samtempa al $[t, P]$ la\u u linio ${\cal L}_a$. Anka\u u supozu ke alia evento $[t_b, Q]$ estas samtempa al $[t, P]$, sed la\u u linio ${\cal L}_b$. Ordinare $t_b\neq t_a$, tial unuigo ${\cal L}_a\cup{\cal L}_b$ ordinare ne estas fermita linio de samtempaj eventoj.   
\newline \selectlanguage{english}Figure~3: $P$ and $Q$ are points of the spatial frame, joined through lines ${\cal L}_a$ and ${\cal L}_b$ in the frame. Suppose event $[t_a, Q]$ is simultaneous to $[t, P]$ along line ${\cal L}_a$. Suppose another event $[t_b, Q]$ is also simultaneous to $[t, P]$, but along line ${\cal L}_b$. Usually $t_b\neq t_a$, so the union ${\cal L}_a\cup{\cal L}_b$ usually is not a closed line of simultaneous events.}
\label{fig.FigLaLbN} 
\end{figure} 

\ppsection[0.6ex]{Sinkronaj horlo\^goj}{Synchronous clocks}                     \label{sinkro}

\ppln{Kutime ni diras ke du horlo\^goj estas sinkronaj se ili `montras la saman horon'. En Newtona mekaniko kaj en speciala relativeco la koordinathorlo\^goj estas \^ciam sinkronaj, \^car ili estas tiel elektitaj. Sed se ni uzas spacotempajn koordinatojn pli \^generalaj, ni bezonas esti pli precizaj por difini sinkronon.}
\pprn{We commonly say that two clocks are synchronous when they `show the same hour'. In Newtonian mechanics and special relativity the coordinate clocks are always synchronous, because they are so chosen. But if we use spacetime coordinates more general, we must be more precise in defining synchrony.} 

\ppl{Por {\it najbaraj} koordinathorlo\^goj, bona difino estas: en elektita molusko, horlo\^goj de ${x_0}^i$ kaj ${x_0}^i+\dd x^i$ estas {\it sinkronaj en momento} $t_0$ se iliaj montroj $t_0$ estas {\it samtempaj} ($\dd T=0$); t.e., se eventoj $[\,t_0, {x_0}^i]$ kaj $[\,t_0, {x_0}^i+\dd x^i]$ estas semtempaj en tiu molusko. Tial, kondi\^co (\ref{dx0}) por samtempeco de najbaraj eventoj, kune kun $\dd t=0$, oferas la {\it kondi\^con por sinkrono} de najbaraj koordinathorlo\^goj en momento $t_0$\,:}
\ppr{For {\it neighbor} coordinate clocks, a good definition is: in a chosen molusc, the clocks of ${x_0}^i$ and ${x_0}^i+\dd x^i$ are {\it synchronous in moment} $t_0$ if their readings $t_0$ are  {\it simultaneous} ($\dd T=0$); that is, if events $[\,t_0, {x_0}^i]$ and $[\,t_0, {x_0}^i+\dd x^i]$ are simultaneous in that molusc. Then the condition (\ref{dx0}) for simultaneity of neighbor events, together with $\dd t=0$, give the {\it condition for synchrony} of neighbor coordinate clocks in moment $t_0$\,:}

\bea                                                                                \label{sinc}
g_{0i}(t_0, {x_0}^i)\,\dd x^i=0. 
\eea

\ppln{Atentu ke sinkrono ordinare estas momenta, \^car $g_{0i}$ ordinare varias la\u u la tempo.}
\pprn{Note that a synchrony is usually momentary, since $g_{0i}$ usually change along time.} 

\ppl{Por koordinathorlo\^goj {\it finie apartitaj} en spaca reto de molusko, oni proponas {\it sinkronon la\u u linio}. Tial, estu evento $[t_x, x^i]$. Kaj selektu linion ${\cal L}$ en spaca reto, entenante punkton $x^i$. Sekvante marku eventojn $[t_y, y^i]$ samtempajn al evento $[t_x, x^i]$\,, en amba\u u direktoj de linio ${\cal L}$\,. La horlo\^go de $y^i$ estas nomita sinkrona al horlo\^go de $x^i$\,, la\u u linio ${\cal L}$ kaj en momento $t_x$\,, se $t_y=t_x$. Atentu ke la kvanto de tiuj aliaj horlo\^goj povas varii de nulo al nefinio, kaj ke la sinkrono ordinare estas momenta.}
\ppr{For coordinate clocks {\it finitely separated} in the spatial frame, one proposes {\it synchrony along a line}. Thus, consider event $[t_x, x^i]$\,, and choose a line ${\cal L}$ in the spatial frame, containing point $x^i$. Then select consecutively events $[t_y, y^i]$ simultaneous to $[t_x, x^i]$, along both directions of line ${\cal L}$\,. The clock of $y^i$ is said synchronous to the clock of $x^i$\,, along ${\cal L}$ in moment $t_x$\,, if $t_y=t_x$. Note that the quantity of these other clocks can vary from zero to infinity, and that the synchrony is usually momentary.}  

\ppl{Klare ke (\ref{sinc}) veri\^gas en moluskoj kun la tri funkcioj $g_{0i}=0$. Tial en tiuj moluskoj \^ciuj koordinathorlo\^goj estas sinkronaj, la\u u iu ajn linio kaj en iu ajn momento, e\^c se la aliaj metrikaj koeficientoj $g_{00}$ kaj $g_{ij}$ pendas de tempo.}
\ppr{Clearly (\ref{sinc}) is true in moluscs with the three functions $g_{0i}=0$. So in these moluscs all coordinate clocks are synchronous, along any line and at any moment, even if the other metrical coefficients $g_{00}$ and $g_{ij}$ depend on time.}

\ppl{Sekcio\,8 montris ke moluskoj havantaj $g_{0i}(x^\mu)/g_{00}(x^\mu)=$ $-c\nabla_i t(x^j)$ permesas \^gene\-ra\-lan samtempecon. Tio sugestas ke ni faru \^san\^gon de tempa koordinato $t\rightarrow t'=t-t(x^j)$, kiu nuli\^gas miksitajn komponojn $g_{0'i'}$ de nova molusko. Tio estas, la koordinathorlo\^goj de nova molusko estos \^ciam sinkronaj.}
\ppr{Section\,8 has shown that moluscs with $g_{0i}(x^\mu)/g_{00}(x^\mu)=-c\nabla_i t(x^j)$ permit global simultaneity. This suggests us to make change of time coordinate $t\rightarrow t'=t-t(x^j)$\,, which brings to zero the mixed components $g_{0'i'}$ in the new molusc. That is, the coordinate clocks of the new molusc will be eternally synchronous.} 

\ppsection[0.6ex]{Komentoj}{Comments}                                        \label{komentoj}

\ppln{Sekcio 8 montris ke molusko hazarde elektita, por spacotempo anka\u u hazarde elektita, ordinare ne permesas tutan samtempecon. Pli klare, ordinare \^gi ne permesas ekziston de eventoj samtempaj la\u u iu ajn linio. Sed iu ajn evento $E=[t, P]$ havas {\it infiniteziman samtempan najbaron}. Alivorte, \^ciam ekzistas infinitezima najbaro de punkto $P$ en spaca reto, kies koordinathorlo\^goj obeas samtempecon (\ref{dx0}). Oni facile provas ke \^ciu paro de eventoj de tiu najbaro estas anka\u u samtempaj.}
\pprn{Section\,8 showed that a randomly chosen molusc, for a spacetime also randomly chosen, usually does not permit global simultaneity. That is, usually it does not permit existence of events that are simultaneous along any line. But any event $E=[t, P]$ has an {\it infinitesimal simultaneous neighborhood}. In other words, there always exist an infinitesimal neighborhood of point $P$ in the spatial frame, whose coordinate clocks satisfy simultaneity (\ref{dx0}). One easily proves that every pair of events in that neighborhood are also simultaneous.}

\ppl{{\it Infinitezima sinkrona disko} estas interesa subaro de tiu najbaro. \^Gi estas infinitezima disko en spaca reto kies koordinathorlo\^goj estas sinkronaj. Do, \^ciu ajn evento estas centro de tial disko.}
\ppr{The {\it infinitesimal synchronous disc} is an interesting subset of that neighborhood. It is an infinitesimal disc in the spatial frame, whose coordinate clocks are synchronous. Thus, every event is center of one such disc.}   

\ppl{{\it Sinkrona linio} en spaca reto ekzistas en molusko kie unu el la tri komponoj $g_{0i}$ estas nula. Ekzemple, se $g_{01}=0$\,, tial (\ref{dx0}) veri\^gas en linio kie nur $x^1$ varias. Simile, {\it sinkrona surfaco} en spaca reto ekzistas en molusko kie du el la tri funkcioj $g_{0i}$ estas nulaj. Fakte, en tia surfaco \^ciuj koordinathorlo\^goj estas sinkronaj la\u u iu ajn linio kunigante ilin, en la surfaco.}
\ppr{A {\it synchronous line} in the spatial frame exists in molusc where one of the three components $g_{0i}$ is null. For example, if $g_{01}=0$\,, then (\ref{dx0}) is true along any line where only $x^1$ varies. Similarly, a {\it synchronous surface} in the spatial frame exists in molusc where two of the three components $g_{0i}$ are null. Remember, in that surface all coordinate clocks are synchronous along any line connecting them, in the surface.}

\ppl{Pli \^generale, sinkronaj surfacoj ekzistas en moluskoj kun}
\ppr{More generally, synchronous surfaces exist in moluscs with}   

\bea                                                                              \label{sinc6}
g_{0i}(t,x^j)=\varphi(t,x^j)\frac{\partial\phi(x^j)}{\partial x^i}\,; 
\eea 
\vskip-2mm 

\ppln{en tiuj moluskoj, surfacoj $\phi(x^j)=$ konst estas sinkronaj.}
\pprn{in these moluscs, the surfaces $\phi(x^j)=$ const are synchronous.}




\ppsection[0.6ex]{Gravita Doppplera \\ efiko}{Gravitational Doppler \\ effect}      \label{dopgrav}

\ppln{En onda fenomeno, Dopplera efiko estas \^san\^go de frekvenco $\nu_{obs}$ observata, kompare kun frekvenco $\nu_{fon}$ igita el fonto. Tiu fenomeno estas priskribita per Dopplera faktoro $D:=\nu_{obs}/\nu_{fon}$\,. Se $D>1$\,, oni diras ke alblui\^go okazas, kaj se $D<1$\,, oni diras ke alru\^gi\^go okazas. Kelkaj kialoj por la neegaleco $D\neq1$ estas rapido kaj loko de fonto en momento de eligo de ondo, kaj anka\u u rapido kaj loko de observanto en momento de ricevo.}
\pprn{In a wave phenomenon, Doppler effect is a change in the observed frequency $\nu_{obs}$\,, compared with the emitted frequency $\nu_{fon}$\,. This effect is measured by the Doppler factor $D:=\nu_{obs}/\nu_{fon}$\,. If $D>1$\,, one says that blueshift occurred, and if $D<1$\,, one says that redshift occurred. Some reasons for the shifts are the velocity and localization of the emitting source in the moment of emission, and also of the observer in the moment of reception.}  

\ppl{Praktika maniero por scii\^gi Doppleran faktoron estas unue supozi ke fonto eligas du signalojn, apartigitajn kun sufi\^ce malgranda propra intertempo $\Delta\tau_{fon}$, mezurita per fonto; poste kalkuli la propran intertempon $\Delta\tau_{obs}$ mezurita per observanto, inter ricevo de tiuj signaloj. Tial la Dopplera faktoro estos~\cite{rdecuamII}}
\ppr{A practical way of obtaining the Doppler factor is first suppose that the source emits two signals, separated by a sufficiently short proper intertime $\Delta\tau_{fon}$ of the source; then calculate the proper intertime $\Delta\tau_{obs}$ of the observer, between the reception of the two signals. The Doppler factor then is~\cite{rdecuamII}}  

\bea                                                                                            
D=\Delta\tau_{fon}/\Delta\tau_{obs}\,.   
\eea

\ppl{En speciala okazo ke fonto kaj observanto estas fiksitaj en spaca reto de nemovebla molusko (tiu, kiu ne malformi\^gas la\u u la tempo), tial~\cite[pa\^go$\,$414]{Anderson}}
\ppr{In the special case that source and observer are fixed in the spatial frame of a stationary molusc (that which does not deform along the time), then~\cite[page$\,$414]{Anderson}} 

\bea                                                                                            
D=\sqrt{\frac{g_{00}(P_{fon})}{g_{00}(P_{obs})}}\,, 
\eea 

\ppln{estante $P_{fon}$ kaj $P_{obs}$ pozicioj de fonto kaj observanto en reto. En molusko de Schwarzschild, ekzemple, kie $g_{00}=1-2Gm/(c^2r)$\,, la Dopplera faktoro por lumo falanta de alto $h_{fon}$ \^gis alto $h_{obs}$ \^ce nia planedo (kun radiuso $R$ kaj maso $M$) estas}
\pprn{being $P_{fon}$ and $P_{obs}$ the locations of source and observer in the frame. In Schwarzschild's molusc, for example, where $g_{00}=1-2Gm/(c^2r)$\,, the Doppler factor for light falling from altitude $h_{fon}$ till altitude $h_{obs}$ in our planet (with radius $R$ and mass $M$) is} 

\bea                                                                                            
D=\sqrt{\frac{1-(2GM/c^2)/(R+h_{fon})}{1-(2GM/c^2)/(R+h_{obs})}}\approx1+\frac{g}{c^2}\Delta h\,, 
\eea   

\ppln{estante $g:=GM/R^{\,2}$ la akcelo pro tera gravito en marnivelo kaj $\Delta h:=h_{fon}-h_{obs}$\,.} 
\pprn{being $g:=GM/R^{\,2}$ the acceleration due to earth gravity at sea-level and $\Delta h:=h_{fon}-h_{obs}$\,.} 

\ppsection[0.6ex]{Reveno al estinto}{Time travel}                            \label{vojagxo}

\ppln{\^Generala relativeco permesas surprizan eblecon, {\it revenon al estinto}. Tio signifas ke persono revenanta de voja\^go povas vidi lokajn horlo\^gojn montrante momenton anta\u uan al de ekvoja\^go. Tiu ebleco estas ofte trovita en moluskoj rotaciantaj rilate al inercia referencosistemo.}
\pprn{General relativity allows for a surprising possibility, {\it time travel}. That is, a person returning from a tour could see the local clocks showing a moment prior to that of departure. That possibility is often found in moluscs rotating relative to an inertial reference system.}   

\ppl{Ekzemple, konsideru spacotempon de Som-Raychaudhuri~\cite{SomRay},}
\ppr{For example, consider the spacetime of Som-Raychaudhuri~\cite{SomRay},}   

\bea                                                                                \label{ds22}
\epsilon(\dd s)^2=[c\,\dd t+(\omega r^2/c)\,\dd\varphi]^2-r^2(\dd\varphi)^2-(\dd r)^2-(\dd z)^2\,. 
\eea

\ppln{La materio kongruanta al tiu metriko estas elektre \^sar\^gita polvo, rigide rotacianta \^cirka\u u akso $z$ de inercia referencosistemo. Sen perdi \^generalecon, ni supozas rotacion en hora direkto ($-\vec{z}$ ), implicante $\omega>0$. La polvo restas en molusko de (\ref{ds22}). Ni montros ke persono revenas (pli maljune) al loka estinto, se voja\^gas en cirklo kun $r>c/\omega$ konstanta, kaj kun konstanta rapido $v>c^2/(\omega r)$\,, en direkto mala al tiu de $\omega$\,.} 
\pprn{The matter related to this metric is electrically charged dust, rotating rigidly around the {\it z}-axis of an inertial reference system. Without loss of generality, we suppose rotation in clockwise direction ($-\vec{z}$ ), implying $\omega>0$. The dust is at rest in the molusc of (\ref{ds22}). We shall show that a person returns (more old) to the local past, if travels in a cirkle with radius $r>c/\omega$\,, and constant velocity $v>c^2/(\omega r)$\,, in direction opposite to $\omega$\,.} 

\ppl{Por nia studo en cirklo, sufi\^cas konsideri}
\ppr{For our study in the cirkle, it suffices consider}
 
\bea                                                                               \label{basta}
(c\,\dd\tau)^2=[c\,\dd t+(\omega r^2/c)\,\dd\varphi]^2-r^2(\dd\varphi)^2\,. 
\eea

\ppln{Komence, atentu ke du najbaraj koordinathorlo\^goj en sama cirklo $r=$ konst ne estas sinkronaj la\u u la cirklo. Fakte, du najbaraj eventoj $[t, r, \varphi]$ kaj $[t+\dd t, r, \varphi+\dd\varphi]$ estas samtempaj se $g_{0\mu}\dd x^\mu=0$\,, t.e., se}
\pprn{To begin, note that two coordinate clocks neighbor in the same cirkle $r=$ const are nor synchronous along the circle. Really, two neighbor events $[t, r, \varphi]$ and $[t+\dd t, r, \varphi+\dd\varphi]$ are simultaneous   if $g_{0\mu}\dd x^\mu=0$\,, that is, if}

\bea                                                                                            
c\,\dd t+(\omega r^2/c)\dd\varphi=0\,. 
\eea

\ppln{Supozante ke $\omega$ kaj $\dd\varphi$ estas pozitivaj, veri\^gas $\dd t<0$\,, indikante ke koordinathorlo\^go de $\varphi+\dd\varphi$ malfruas rilate al tiu de $\varphi$\,. Do, se io ekvoja\^gas el $\varphi_0$ en loka momento $t_0$\,, en malhora direkto (direkto {\it mala} al rotacio $\omega$), kun nefinia rapido, tial \^gi atingos pozicion $\varphi_0+\dd\varphi$ en loka momento $t_0-(\omega r^2/c^2)\dd\varphi$\,. Se tiu io da\u uras kun nefinia rapido en cirklo $r=$ konst, \^gi revenos al komenca pozicio $\varphi_0$ en loka momento $t_0-(2\pi\omega r^2/c^2)$\,. Tio estas, \^gi revenos en momento anta\u ua al tiu de ekvoja\^go.}
\pprn{Supposing $\omega$ and $\dd\varphi$ positive, then $\dd t<0$\,, indicating that the coordinate clock in $\varphi+\dd\varphi$ is late relative to that in $\varphi$\,. Thus, if something starts from $\varphi_0$ in local moment $t_0$\,, in anti-clockwise direction (direction {\it opposite} to the rotation $\omega$), with infinite velocity, it reaches position $\varphi_0+\dd\varphi$ in local moment $t_0-(\omega r^2/c^2)\dd\varphi$\,. If that something proceedes with infinite velocity in the cirkle $r=$ const, it returns to position $\varphi_0$ in local moment $t_0-(2\pi\omega r^2/c^2)$\,. That is, it returns in moment prior to that of departure.} 

\ppl{Sed persono ne povas havi $|v|=\infty$\,, li devas havi $|v|<c$. Tial ni demandas: \^cu estas rapido $|v|<c$ kaj radiuso $r$ tiaj ke momento de reveno de voja\^ganto estas frua je momento de ekvoja\^go, en universo de Som-Raychaudhuri? Surpriza respondo de \^generala relativeco estas {\it jes}.}
\ppr{But a person cannnot have $|v|=\infty$\,, he must have $|v|<c$. We then ask: is there any velocity $|v|<c$ and radius $r$ such that the moment of return of the voyager is prior to the moment of departure, in Som-Raychaudhuri universe? The surprising answer of general relativity is {\it jes}.}

\ppl{Fakte, rapido de voja\^ganto en cirklo estas $v=\dd L/\dd T$\,, estante $\dd L=r\dd\varphi$ la\u u (\ref{dL}), kaj $\dd T=\dd t+(\omega r^2/c^2)\dd\varphi$ la\u u (\ref{dT}). Tial}
\ppr{Really, the velocity of voyager in the circle is $v=\dd L/\dd T$\,, with $\dd L=r\dd\varphi$ according to (\ref{dL}), and $\dd T=\dd t+(\omega r^2/c^2)\dd\varphi$ according to (\ref{dT}). Thus}   

\bea                                                                                \label{vret}
v=\frac{r\dd\varphi}{\dd t+(\omega r^2/c^2)\dd\varphi}\,. 
\eea

\ppln{\^Car $v=$ konst implicas $\dd\varphi/\dd t=\Delta\varphi/\Delta t$\,, tial en kompleta turno ($\Delta\varphi=2\pi$) okazas}
\pprn{Since $v=$ const implies $\dd\varphi/\dd t=\Delta\varphi/\Delta t$\,, in a complete turn  ($\Delta\varphi=2\pi$) it occurs} 

\bea                                                                                            
v=\frac{2\pi r}{\Delta t+2\pi\omega r^2/c^2}\,. 
\eea 

\ppln{Do}
\pprn{Then}  

\bea                                                                                            
\Delta t=2\pi r(1/v-\omega r/c^2)\,, 
\eea 

\ppln{montrante ke $\Delta t$ estas negativa se $v>c^2/(\omega r)$\,. \^Car ni impozas $v<c$, la kondi\^co por persono reveni al estinto en spacotempo de Som-Raychaudhuri estas}
\pprn{showing that $\Delta t$ is negative whenever $v>c^2/(\omega r)$\,. Since we impose $v<c$, the condition to a person return to the past in the spacetime of Som-Raychaudhuri is} 

\bea                                                                                \label{kond}
c/(\omega r)<v/c<1\,. 
\eea 

\ppln{Figuro~\ref{FigRetorno} indikas regionon de ebeno $[r, v]$ rilata al reveno de voja\^ganto al estinto.}
\pprn{Figure~\ref{FigRetorno} indicates the region of the plane $[r, v]$ related to return of a voyager to the past.}

\begin{figure}                                                                            
\centerline{\epsfig{file=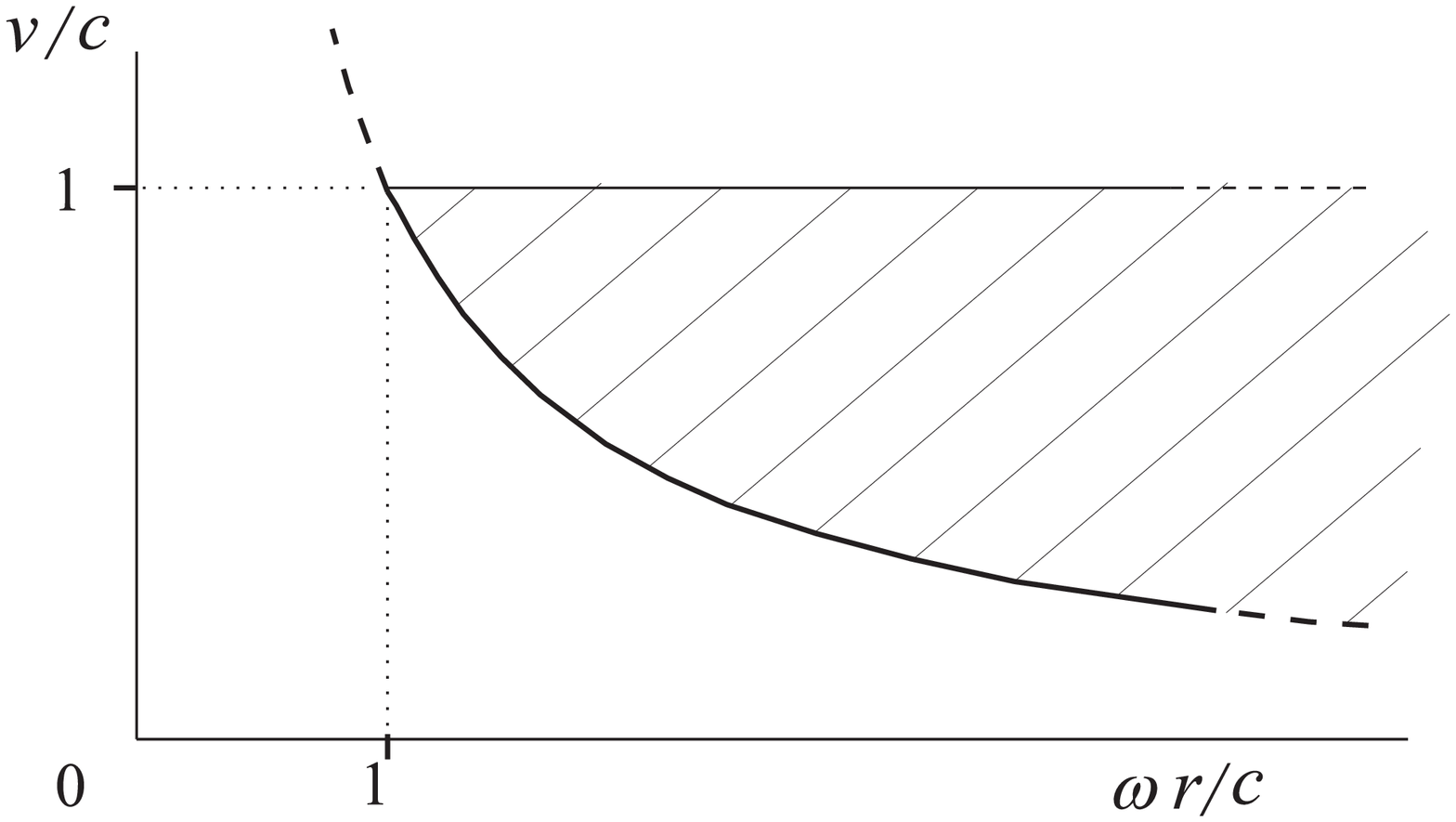,width=8cm}} 
\selectlanguage{esperanto}\caption{\selectlanguage{esperanto} Strekata regiono, $c/(\omega r)<v/c<1$, indikas parojn de radiuso $r$ (kun unueco $c/\omega$) kaj rapido $v$ (kun unueco $c$) rilataj al reveno de voja\^ganto al estinto, en spacotempo de Som-Raychaudhuri.   
\newline \selectlanguage{english}Figure~4: The streaked region, $c/(\omega r)<v/c<1$, indicates pairs of radius $r$ (with unit $c/\omega$) and velocity $v$ (with unit $c$) related to return of a voyager to the past, in spacetime of Som-Raychaudhuri.}
\label{FigRetorno} 
\end{figure} 


\ppsection[0.6ex]{Konkludo}{Conclusion}

\ppln{Pluraj niaj fruaj artikloj diskutis pri spacotempo \^ce la speciala relativeco~\cite{reltemp1,rdecuamII,pluraj}. Aliaj artikloj traktis pri fenomenoj en \^generala relativeco~\cite{GeSiSchw,SomRay,ppluraj}. Ni nun diskutis plidetale pri spacotempo \^ce la \^generala relativeco.}
\pprn{Several preceding articles of ours discoursed upon spacetime in the special relativity~\cite{reltemp1,rdecuamII,pluraj}. Other articles dealt with phenomena in the general relativity~\cite{GeSiSchw,SomRay,ppluraj}. We now discoursed with more details upon spacetime in the general relativity.}

\ppl{Sekcio 1 traktis pri temoj gravaj por la artiklo: normhorlo\^go, propratempo, konstanteco de $c$\,, kaj sekondo.}
\ppr{Section 1 dealt with important topics for the article: standard clock, propertime, constancy of $c$\,, and second.}

\ppl{Sekcio 2 emfazis kiom libere la spacotempaj koordinatoj estas elektitaj en \^generala relativeco, speciale la loka tempa koordinato. Komencantaj fizikistoj ofte ne kredas ke tiel \^generalaj koordinatoj povas esti utilaj por precizaj kalkuloj.}
\ppr{Section 2 emphazised how freely the spacetime coordinates are chosen in general relativity, especially the local time coordinate. Pedestrian physicists often do not believe that so general coordinates can be useful for precise calculations.}

\ppl{Sekcio 3 montris kiel la spacotempa metriko rilatas tiujn arbitrajn koordinatojn al infinitezimaj propraj distancoj $\dd\lambda$\,, a\u u al infinitezimajn propraj intertempoj $\dd\tau$. Por tio, sole la konstanta rapido $c$ de lumo en vakuo estis uzita.}
\ppr{Section 3 showed how the spacetime metric relates these arbitrary coordinates to infinitesimal proper distances $\dd\lambda$\,, or to infinitesimal proper intertimes $\dd\tau$. To that end, solely the constant velocity $c$ of light in vacuum was used.}

\ppl{Sekcioj 4 -- 6 difinis distancon $\dd L$ inter du najbaraj punktoj fiksitaj en spaca reto, kaj intertempon $\dd T$ de evento al najbara evento. Amba\u u $\dd L$ kaj $\dd T$ estis difinitaj en inercia referencosistemo restanta rilate al elektita spaca reto, najbare al la eventoj. Oni ordinare konsentas ke $\dd L$ kaj $\dd T$ estas kvantoj kiuj plej similas al Newtona distanco kaj intertempo. Tri-dimensia metriko $h_{ij}(x^\mu)$ por la spaca reto estis derivita, denove uzante sole la konstantan valoron $c$\,. Plu, rapido $v$ rilatante du najbarajn eventojn estis prezentita, pendanta de spacotempa koordinataro uzita. \^Gia modulo superas $c$\,, se intervalo estas de spaca tipo.}
\ppr{Sections 4 -- 6 defined distance $\dd L$ between two neighbor points fixed in the spatial frame, and intertime $\dd T$ from one event to a neighbor one. Both $\dd L$ and $\dd T$ were defined in an inertial reference system at rest relative to the local frame under use, in the vicinity of the events. One usually agrees that $\dd L$ and $\dd T$ are the quantities that most closely resemble the Newtonian distance and intertime. A three-dimensional metric $h_{ij}(x^\mu)$ for the spatial frame was derived, again using solely the constant value $c$\,. Still, a velocity $v$ relating two neighbor events was presented, depending on the spacetime coordinates used. Its modulus exceeds $c$\,, if the interval is spacelike.}

\ppl{Sekcio 7 klarigis ke, en \^generala relativeco, distancoj kaj intertempoj rilataj al eventoj finiaj apartitaj ne estas kiel en Newtona mekaniko. Nun bezonas enkonduki koncepton de distanco kaj de intertempo la\u u elektita kurbo en spacotempo. Simile, Sekcioj 8 kaj 9 difinis samtempecon de eventoj kaj sinkronon de horlo\^goj, la\u u elektita linio en spaca reto.}
\ppr{Section 7 made clear that, in general relativity, distances and intertimes relating events finitely separated are not as in Newtonian mechanics. Now needs introduce the concept of distance and intertime along a chosen curve in spacetime. Similarly, Sections 8 and 9 defined simultaneity of events and synchrony of clocks, along a chosen line in the spatial frame.}

\ppl{Sekcio 10 anoncis ekziston de du interesaj infinitezimaj subaroj de spaca reto. La unua estas tri-dimensia, permesante ekziston de samtempaj eventoj. La dua estas du-dimensia, entenante sinkronajn koordinathorlo\^gojn.}
\ppr{Section 10 announced existence two interesting infinitesimal subsets of the spatial frame. The first is three-dimensional, permitting simultaneous events. The second is two-dimensional, containing synchronous coordinate clocks.}

\ppl{Sekcio 11 prezentis gravan prognozon de \^generala relativeco, tion de gravita alblui\^go de lumo falanta sur tero. Alia prognozo estas gravita alru\^gi\^go de lumo eskapanta el stelo. Kutime tiuj du efikoj konkursas, kaj la gravita alru\^gi\^go ordinare estras.}
\ppr{Section 11 presented an important prediction of general relativity, the gravitational blueshift of light falling on earth. Another prediction is the gravitational redshift of light escaping from a star. In \^general these two effects conflict, and the redshift usually dominates.}

\ppl{Fine, Sekcio 12 faris detalan eksponadon de reveno al estinto en spacotempo de Som-Raychaudhuri. En tiu universo kun $\omega=$ 1 turno/jarcento, persono revenas al estinto se voja\^gas en cirklo kun radiuso $r=$ 70 lumjaroj kun konstanta rapido $v=c/4$\,, en direkto mala al $\omega$\,. Kruda takso prognozas komfortan radiusan akcelon (\^cirka\u u trionon de tera gravito \^ce marnivelo), por teni voja\^ganton en orbito.}
\ppr{Finally, Section 12 made a detailed exposition of time travel in the Som-Raychaudhuri spacetime. In such universe with $\omega=$ 1 turn/century, a person returns to the past, if travels in a circle with radius $r=$ 70 light-years with constant velocity $v=c/4$\,, in direction opposite to $\omega$\,. A crude estimate predicts a comfortable radial acceleration (nearly one third of earth gravity at sea-level), to maintain the voyager in orbit.}

\selectlanguage{esperanto}

\end{Parallel}


\begin{thebibliography}{100}
\bibitem{tipos} Malsimile al $\Delta\tau$ kaj $\Delta\lambda$\,, amba\u u $\Delta T$ kaj $\Delta L$ estas difineblaj en kurboj kun pecoj de malsimilaj tipoj.  
\\ Differently from $\Delta\tau$ and $\Delta\lambda$\,, both $\Delta T$ and $\Delta L$ can be defined in curves with pieces of different types. 
\end{thebibliography}
\end{document}